\documentclass[useAMS,usenatbib,usegraphicx]{mn2e}
\usepackage{subfigure}
\usepackage{mathabx}

\newcommand{\cs}{c_{\rm s}}
\newcommand{\dvl}{\delta v_{\ell}}
\newcommand{\Eg}{E_{\rm g}}
\newcommand{\Ek}{E_{\rm k}}
\newcommand{\Jeff}{J_{\rm eff}}
\newcommand{\kms}{{\rm ~km~s}^{-1}}
\newcommand{\kpk}{k_{\rm pk}}
\newcommand{\LJ}{L_{\rm J}}
\newcommand{\ls}{\lambda_{\rm s}}
\newcommand{\Ma}{{\cal M}_{\rm A}}
\newcommand{\Ms}{{\cal M}_{\rm s}}
\newcommand{\pcc}{{\rm ~cm}^{-3}}
\newcommand{\psc}{{\rm ~cm}^{-2}}
\newcommand{\Rjeff}{R_{\rm J,mt}}
\newcommand{\sfeff}{{\rm SFE}_{\rm ff}}
\newcommand{\tff}{t_{\rm ff}}
\newcommand{\tgrav}{t_{\rm grav}}
\newcommand{\tturb}{t_{\rm turb}}
\newcommand{\va}{v_{\rm A}}
\newcommand{\vv}{{\bf v}}
\newcommand{\vrms}{v_{\rm rms}}

\newcommand{\aap}{A\&A}
\newcommand{\aaps}{A\&AS}

\newcommand{\apj}{ApJ}

\newcommand{\apjs}{ApJS}

\newcommand{\araa}{ARAA}
\newcommand{\mnras}{MNRAS}
\newcommand{\pasj}{PASJ}

\newcommand{\pre}{Phys.\ Rev.\ E}

\newcommand{\beq}{\begin{equation}}
\newcommand{\eeq}{\end{equation}}
\newcommand{\bea}{\begin{eqnarray}}
\newcommand{\eea}{\end{eqnarray}}

\title[Testing star formation theories]{Testing
assumptions and predictions of star formation theories}
\author[ Alejandro Gonz\'alez-Samaniego
  et al.]{Alejandro Gonz\'alez-Samaniego$^{1}$\thanks{ e-mail: ags@astro.unam.mx},
Enrique V\'azquez-Semadeni$^{1}$, \newauthor
Ricardo F. Gonz\'alez$^{1}$,
and Jongsoo Kim$^{2}$\\
$^{1}$Centro de Radioastronom\'ia y Astrof\'isica, Universidad
  Nacional Aut\'onoma de M\'exico, Apdo. Postal 3-72, Morelia, 58089, M\'exico \\
$^{2}$Korea Astronomy and Space Science Institute, 61-1, Hwaam-dong,
  Yuseong-gu, Daejon 305-764, Korea\\
\\
$\star$ Present address: Instituto de Astronom\'{\i}a, Universidad Nacional Aut\'onoma de M\'exico, 
A.P. 70-264, 04510, M\'exico, D.F., M\'exico}

\begin{document}

\maketitle
\label{firstpage}
\begin{abstract}
We present numerical simulations of isothermal, magnetohydrodynamic (MHD), supersonic
turbulence, designed to test various hypotheses frequently assumed in
star formation (SF) theories. This study complements our previous one in
the non-magnetic (HD) case. We consider three simulations, each with
different values of its physical size, rms sonic Mach number $\Ms$, and
Jeans parameter $J$, but so that all three have the same value of the
virial parameter and conform with Larson's scaling relations. As in the
non-magnetic case, we find that (1) no structures that are both subsonic
and super-Jeans are produced; (2) that the fraction of small-scale
super-Jeans structures increases when self-gravity is turned on, and the
production of very dense cores by turbulence alone is very low. This
implies that self-gravity is involved not only in the collapse of
Jeans-unstable cores, but also in their formation. (3) We also find that
denser regions tend to have a stronger velocity convergence, implying a
net inwards flow towards the regions' centres. Contrary to the
non-magnetic case, we find that the magnetic simulation with lowest
values of $\Ms$ and $J$ (respectively, 5 and 2) does not produce any
collapsing regions for over three simulation free-fall times, in spite
of being both Jeans-unstable and magnetically supercritical. We
attribute this result to the combined thermal and magnetic support.
Next, we compare the results of our HD and MHD simulations with the
predictions from the recent SF theories by Krumholz \& McKee,
Padoan \& Nordlund, and Hennebelle \& Chabrier, using
expressions recently provided by Federrath \& Klessen, which extend
those theories to the magnetic case. In both the HD and MHD cases, we
find that the theoretical predictions tend to be larger than the
$\sfeff$ measured in the simulations. In the MHD case, none of the
theories captures the suppression of collapse at low values of $\Jeff$
by the additional support from the magnetic field. We conclude that
randomly driven isothermal turbulence may not correctly represent the
flow within actual clouds, and that theories that assume this regime may
be missing a fundamental aspect of the flow. Finally, we suggest that a
more realistic regime may be that of hierarchical gravitational
collapse.

\end{abstract}

\begin{keywords}
turbulence -- magnetic fields -- ISM: clouds -- stars: formation. 
\end{keywords}

\section{Introduction} \label{sec:intro}

Theories of star formation (SF) necessarily rely on assumptions about the
structure of the molecular clouds (MCs) where the stars form. One of the
most common of such assumptions is that the observed supersonic
linewidths observed within MCs constitute supersonic turbulence
\citep{ZE74}, and that this turbulence causes a net pressure that
provides support against the clouds' self-gravity, maintaining them in
approximate virial equilibrium \citep[e.g.,][]{Larson81, MG88,
Blitz93}. The notion of turbulent pressure as an agent of support
against self-gravity was first introduced by \citet{Chandra51} and later
the scale-dependence of the turbulent support was also considered
\citep[e.g.,][]{Bonazzola+87, VG95}.

It is important to note that, in order for the non-thermal motions to be
able to provide support against self-gravity, the turbulent velocity
field must be essentially random. This property is actually extremely
difficult to measure observationally, because one normally has
information about only {\it one} component of the vector velocity field
(the one along the line of sight), and moreover one does not have
information about the spatial structure of that component along the line
of sight. Thus, most measurements of the velocity structure in MCs refer
to the {\it magnitude} of the velocity, which in turn is most commonly
interpreted in terms of the non-thermal kinetic energy in the clouds, and
its relative importance in the clouds' virial balance, in particular in
relation to the clouds' gravitational energy \citep[e.g.,] [and
references therein]{MK04, MO07}. Statistical quantities, such as the
kinetic energy {\it spectrum} in MCs \citep[see, e.g., the review
by][]{ES04}, give the distribution of the kinetic energy over the range
of size scales present in the turbulent motions. However, this
distribution is still obtained by averaging over the cloud's volume, so
no spatial information on the velocity field structure is provided by
it.

Studies of the radial velocity gradient
\citep[e.g.][]{GA85, Goodman+93, Phillips99, Caselli+02, Rosol07, IB11,
IBB11, Li+12} provide a crude approximation to the spatial structure of
the radial component of the velocity field across MCs. It must be noted
that such gradients are most often interpreted as rotation, although
in principle they might equally be interpreted as shear, expansion, or
contraction. In fact, \citet{Brunt03} has pointed out that a principal
component analysis of the clouds is inconsistent with that
expected for uniform rotation, and instead interprets the inferred
principal components of the velocity dispersion as large-scale chaotic
turbulence \citep[see also][]{Brunt+09}. Moreover, the possibility that
the clouds are undergoing global contraction has been advocated by
various recent studies \citep{VS+07, VS+09, HH08, BP+11a, BP+11b}, in
which case the gradients might be interpreted as the signature of this
contraction. Thus, determination of the velocity field's topology in MCs
is a very important challenge for our understanding of their structure.

Another universal notion concerning SF is that, in MCs and their
substructure, there exists a linewidth--size relation of the form
\citep{Larson81}
\beq
 \dvl \propto  \ell ^ \kappa,
\label{eq:dv-r}
\eeq
and it is widely believed that such scaling relation originates as a
consequence of the existence of a turbulent cascade, which produces a
kinetic energy spectrum of the form $E(k) \propto k^{-n}$. Indeed, this
form of the energy spectrum implies that the typical velocity difference
across scales $\ell \sim 1/k$ is $\dvl \propto \ell^{(n-1)/2}$
\citep[see, e.g.,][]{VS+00}.  Thus, a spectral slope $n=5/3$ directly
implies $\dvl \propto \ell^{1/3}$, close to the value $\kappa \approx
0.38$ initially observed by \citet{Larson81}, while $n=2$ implies $\dvl
\propto \ell^{1/2}$, as determined by most later studies
\citep[see, e.g., the review by][and references therein]{Blitz93}. 

\citet{Larson81} also advanced a scaling relation between mean density
and size, of the form 
\beq
\langle \rho \rangle \propto \ell ^ \lambda,
\label{eq:n-r}
\eeq
with $\lambda \sim -1$. Relations (\ref{eq:dv-r}) and (\ref{eq:n-r}) are
almost universally thought to be a manifestation of near virial
equilibrium in the clouds, as originally proposed by \citet{Larson81}
himself. However, the density--size relation has been questioned by a
number of authors \citep[e.g.,][]{Kegel89, Scalo90}, in particular as it
may be the result of an observational selection effect. Moreover,
\citet{BP+11a} have recently pointed out that the linewidth--size
relation is in general not universally verified, and that it might also
be a consequence of the same selection effects. Indeed, high-mass
star-forming clumps typically exhibit larger linewidths for their size
than those implied by the
\citet{Larson81} relation \citep[e.g.,][]{CM95, Plume+97, Shirley+03,
Gibson+09, Wu+10}. Yet, \citet{BP+11a} showed that, when the massive
cores have mass determinations independent from virial estimates, it is
found that their linewidth, size, and column density follow the
general relation found by \citet{Heyer+09}, namely 
\beq
 \dvl\propto \Sigma^{1/2} \ell^{1/2}.
\label{eq:dv-r-Sigma}
\eeq
The latter authors interpreted this
scaling as evidence of virial equilibrium in clouds without constant
column density, although \citet{BP+11a} noted that this may just as well
be interpreted as evidence of free-fall in the clouds and their
substructures.

Within the context of the linewidth--size relation, equation (\ref{eq:dv-r}),
it is well known that such a relation implies that at some particular
scale, termed the `sonic scale' ($\ls$), the turbulent velocity
dispersion is equal, on average, to the sound speed in the medium. Thus,
a number of authors \citep[e.g.,][]{Padoan95, VBK03, KM05, Federrath+10}
have assumed that stars form in cores that have sizes equal to or smaller
than the sonic scale, if they contain more than one thermal Jeans mass,
since in these cores the main support is thermal. In principle, such
cores can proceed to collapse without significant turbulent support nor
fragmentation. Clumps larger than this scale are assumed to be
globally supported
against their self-gravity by turbulence, since they exhibit scaling
relations like relation (\ref{eq:dv-r-Sigma}), and thus are interpreted
as being near virial equilibrium. Nevertheless, within the clumps, the
turbulence is assumed to induce local compressions that constitute
smaller scale clumps and cores. However, if MCs are in a global state
of gravitational collapse, as proposed by \citet{VS+07, VS+09},
\citet{HH08} and \citet{BP+11a}, then the observed linewidths may not
reflect supporting turbulent motions, but rather the infall velocities
themselves. In this case, the notion that the structures that collapse
are the subsonic, super-Jeans cores may not apply.


In a previous paper \citep[][hereafter Paper I]{VS+08}, we presented a
numerical study designed to test the hypotheses mentioned above, using
simulations of supersonic, isothermal, hydrodynamic driven
turbulence. We considered three simulations with different physical box
sizes, rms Mach numbers ($\Ms$), and Jeans numbers ($J = L/\LJ$, with
$L$ the numerical box size and $\LJ$ the Jeans length), but such that
the ratio $\Ms/\LJ$ remained constant, thus assuring that the boxes
followed Larson-type scaling relations \citep{Larson81}. We found, among
other results, that (a) there appeared to exist a negative correlation
between the mean density and the mean velocity divergence of isolated
subregions in the flow, suggesting that the velocity field is not
completely random in overdense regions, but is characterized by a net
convergence (negative divergence) of the velocity. The fact that the
flow within the clumps has a net negative divergence instead of being
fully random implies that not all of the flow's kinetic energy is
available for support against the self-gravity of the clump.
(b) Clumps or
subboxes of the numerical box with subsonic velocity tended to be Jeans
{\it stable}, although significant gravitational collapse did occur in
the simulations. This suggested that the main collapsing structures are
large-scale, supersonic clumps, rather than small-scale, subsonic ones.
(c) The SF efficiency per free-fall time of the various
simulations, $\sfeff$, scaled with $\Ms$ roughly in agreement with the
theoretical prediction by \citet[][hereafter KM05]{KM05}, within the
(relatively large) uncertainties. 

Since the publication of Paper I, two new theories have
appeared \citep{HC11, PN11}, in addition to that by KM05, which attempt
to describe the dependence of the star formation rate (SFR) on the main
physical parameters of the clouds, namely the ratio of kinetic to
gravitational energy, characterized by the {\it virial parameter}
$\alpha$ (KM05), the rms turbulent Mach number $\Ms$, the Alfv\'enic
Mach number $\Ma$, and the ratio of solenoidal to compressive modes
injected to the turbulence, measured by the so-called {\it b-parameter}
\citep{Federrath+08}. All of those theories
start from considering the fraction of the mass in a turbulent cloud
above a certain critical density, as computed from the probability
density function (PDF) of the density field, expected to have a
lognormal form \citep{VS94, Padoan+97, PV98}. The main difference
between the theories resides mainly in how this `star-forming
fraction' is selected. Specifically, KM05 assumed that stars form from
clumps that simultaneously satisfy the conditions of being {\it
subsonically turbulent} (i.e. have sizes below the sonic scale) and of
having a density large enough that their Jeans length is equal to or
smaller than the sonic scale, as described above. \citet[][hereafter
PN11] {PN11} further included magnetic support to compute the
appropriate density cutoff, while \citet[][hereafter HC11] {HC11} took a
scale-dependent `turbulent support' into account, as well as the fact
that material at different densities evolve on different time-scales,
given by their corresponding free-fall times. A detailed review on how
each theory determines this fraction is provided by \citet[][hereafter
FK12] {FK12}.

It can thus be seen that all three of these recent theories indeed rely
on one of the fundamental assumptions discussed above, namely that the
supersonic non-thermal motions constitute isotropic turbulence that,
while locally inducing compressions that can become Jeans-unstable and
collapse, globally produce a turbulent pressure that can oppose the
clumps' self-gravity. Moreover, two of these theories (KM05 and PN11)
rely on the assumption that simultaneously subsonic and super-Jeans
cores play a fundamental role in the process of SF. Both of
these assumptions were questioned, in the non-magnetic case, in Paper
I. It is thus important to test whether these assumptions are verified,
at least in numerical simulations designed for that purpose. It is worth
pointing out that one theory not assuming turbulent support, but rather
generalized gravitational collapse, has been presented by \citet{ZVC12}.

The assumption that the non-thermal motions in MCs consist of
turbulence that can oppose self-gravity extends beyond theories for the
SFR. In particular, theories for the mass spectrum of the clouds
themselves as well as the dense cores within them have often relied on
this assumption. For example, \citet{HC08} developed a theory for the
stellar initial mass function (actually, for the core mass function in
MCs) that relied on the competition between turbulent support and
self-gravity, and, more recently, \citet{Hopkins12a, Hopkins12b,
Hopkins13} has combined this with an excursion-set formalism in order to
obtain the mass function of gravitationally bound objects (with respect
to thermal, turbulent and rotational support) both at large and small
scales.

In this paper, we continue the study performed in Paper I,
now in the magnetohydrodynamic (MHD) case, aimed at testing the hypotheses that the bulk
motions in the clumps can provide support against self-gravity, and that
clumps that are simultaneously subsonic and super-Jeans are produced by
turbulent compressions. We also use our driven-turbulence simulations to
test the predictions of the various theories for the SFR. The plan of
the paper is as follows: In Sec.\ \ref{sec:models}, we discuss the
control parameters for our numerical simulations and the cases we have
considered. Next, in Sec.\ \ref{sec:num_sim}, we present the results
from the simulations, and in Sec.\ \ref{sec:discussion}, we discuss them
in the context of previous results, including our non-magnetic ones, as
well as their implications. Finally, in Sec.\ \ref{sec:conclusions}, we
present a summary and some conclusions.

\section{The models} \label{sec:models}

\subsection{Control parameters} \label{sec:control_params}

Our simulations of supersonic, isothermal, magnetized, and
self-gravitating turbulence may be described in terms of three
dimensionless parameters, namely the rms turbulent Mach number
$\Ms$, the Jeans number $J$, and the mass-to-magnetic flux ratio (in
units of its critical value), $\mu$. In this paper, we employ the same
normalization as in \citet{VS+05}. The rms turbulent Mach number is
given by $\Ms = \vrms/\cs$, where $\vrms$ is the rms turbulent velocity
dispersion and $\cs$ is the isothermal sound speed. The Jeans number is
defined by $J= L/L_{\rm J}$, where $L$ is the numerical box size and
\beq
L_{\rm J} = \left(\frac{\mathrm{\pi} \cs^2}{G \rho}\right)^{1/2}
\label{eq:LJ}
\eeq
is the Jeans length at density $\rho$ and isothermal sound speed
$\cs$. The magnetic flux $\phi$ is defined as
\beq
\phi \equiv \int_A \textbf{B} \cdot d\textbf{S},
\label{eq:flux}
\eeq
where $A$ is a cross-sectional area across the region over which the
flux is to be evaluated. The critical value of the mass-to-magnetic
flux ratio for a cylindrical geometry
is given by \citep{NN78}
\beq
\left(\frac{M}{\phi}\right)_{\rm crit} = \frac{1}{\sqrt{4 \mathrm{\pi}^2 G}}  \approx
0.16~ G^{-1/2}.
\label{eq:NN78}
\eeq
This is the relevant criterion for our Cartesian simulations, since the
column density is the same along all flux tubes in the initial
conditions, as in a cylindrical configuration. A spherical criterion,
such as that given by \citet{Shu92}, would apply for a configuration in
which the column density is lower for flux tubes intersecting a spherical
cloud near its poles.

It is important to note that collapsing clouds must have $J> 1$
and $\mu> 1$, while clumps with $J> 1$ and $\mu< 1$ are gravitationally
bound but will only contract for a while, and then oscillate around a
stable magnetostatic state
\citep{VS+11}.  Finally, structures with $J< 1$ are Jeans-stable and
must re-expand after being formed by a transient turbulent compression,
regardless the value of $\mu$.

\begin{table*}
 \centering
 \begin{minipage}{320mm}
  \caption{Run parameters.}
  \begin{tabular}{@{}cccccccccccccccc@{}}
  \hline
   Name     &   $L$   &   $n_{0}$   &   M   &   $\LJ$  &   $J$  &   $\tff$   &
   $\tgrav$   &   $\beta$    &  $\va$ &
$\vrms$ & $\Jeff$ & Resolution & Driving \\
            & (pc) &    (cm$^{-3}$)   &   (M$_{\Sun}$)  &   (pc)  &   &  (Myr)  &
   (Myr)   &   &   &   (km s$^{-1}$) &  (km s$^{-1}$)
  \\
 \hline
 Ms5J2  & 1 & 2000   & 115.8 & 0.5 & 2  & 2.5 & 2 &  0.086 
  & 0.967 & 1. & 0.91 & 512 & Sol.\\
 Ms10J4  & 4 & 500    & 1853  & 1  &  4   & 5   & 4 &  0.021 
 & 1.934 & 2. & 0.95 & 512 & Sol.\\
 Ms15J6  & 9 & 222.22 & 9382  & 1.5 &  6  & 7.5 & 6 &  0.0095 
 & 2.901 & 3. & 1.02 & 512 & Sol.\\
 Ms15J6C-128  & 9 & 222.22 & 9382  & 1.5 &  6  & 7.5 & 6 &  0.0095 
 & 2.901 & 3. & 1.02 & 128 & Comp.\\
 Ms15J6C-256  & 9 & 222.22 & 9382  & 1.5 &  6  & 7.5 & 6 &  0.0095 
 & 2.901 & 3. & 1.02 & 256 & Comp.\\
 Ms15J6C-512  & 9 & 222.22 & 9382  & 1.5 &  6  & 7.5 & 6 &  0.0095 
 & 2.901 & 3. & 1.02 & 512 & Comp.\\

\hline
\label{tab:run_parameters}
\end{tabular}
\end{minipage}
\end{table*}

Two other frequently used parameters for describing a magnetized plasma are
the so-called `plasma $\beta$' and the Alfv\'enic Mach number,
$\Ma \equiv \vrms/\va$. The former is given by $\beta= P_{\rm th}/P_{\rm
mag}$, where $P_{\rm th} = \cs^2 \rho$ and $P_{\rm mag} = B^2/8 \mathrm{\pi}$. It
can be easily shown that, for a cubic numerical box of size $L$, and for
uniform initial density and magnetic fields, $\beta$ is related to the
normalized mass-to-flux ratio and the Jeans number by

\begin{equation}
\beta = \frac{2}{\mathrm{\pi}^2} \left( \frac{\mu}{J} \right)^2,
\label{lambda}
\end{equation}
where we have assumed that the critical mass-to-flux ratio is given by
the cylindrical expression, equation (\ref{eq:NN78}). Similarly, the
Alfv\'enic Mach number is related to the nondimensional parameters by
\beq
\Ma = \sqrt{\frac{\beta}{2}} \Ms = \frac{\mu \Ms}{\mathrm{\pi} J}.
\label{eq:M_alfv}
\eeq
This implies that all three of our simulations have $\Ma \approx 1.03$.

Also, from equation (\ref{lambda}), we see that the critical value of
$\beta$ (that is, the value of $\beta$ that corresponds to $\mu =1$) is
$\beta_{\rm cr}= 2/(\mathrm{\pi}^2 J^2)$, and thus we have, in general,
\beq
\mu= (\beta/\beta_{\rm cr})^{1/2}.
\label{eq:mu-beta_rel}
\eeq

In addition, we assume the same magnitude of the magnetic field in all
our runs. This is motivated by the observation that the magnetic
field strength is roughly independent of density for densities below
$\sim 10^3 \pcc$, with magnitudes of a few tens of $\mu$G \citep[see
fig.\ 1 of][]{Crutcher+10}. 

A separate argument is the following. Consider the
definition of the Alfven speed $\va$, that is,
\begin{equation}
\va^2 = {{B_0^2}\over{4\mathrm{\pi} \rho}},
\label{b0}
\end{equation}
which, in terms of the Alfvenic Mach number $\Ma$, reads

\begin{equation}
B_0^2= 4\mathrm{\pi} \rho \biggl({{\Delta v}\over{\Ma}}\biggr)^2\,,
\label{b01}
\end{equation}
where $\Delta v$ is the three-dimensional velocity dispersion.  Then,
inserting the Larson relations (\ref{eq:dv-r}) and (\ref{eq:n-r}), we
see that $B_0$, which is proportional to $\rho\, \Delta_v^2 =$ cst., should be
taken constant in our constant-$\Ma$ numerical simulations.

\subsection{Numerical simulations} \label{sec:num_sim}

\begin{figure}
\includegraphics[scale=0.45]{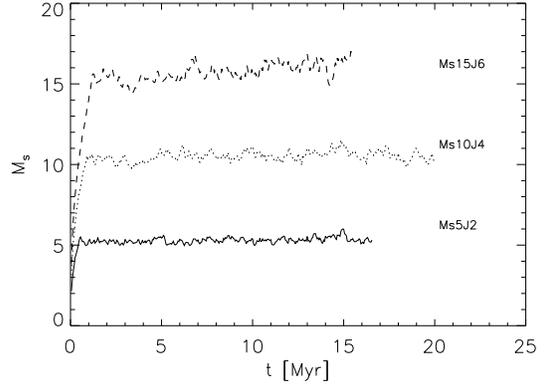}
\caption{Evolution of the turbulent Mach number $\Ms$ for each of our
simulations.}
\label{fig:vrms_vs_t}
\end{figure}

As mentioned in Sec.\ \ref{sec:intro}, we consider a suite of three main
numerical simulations of randomly driven, isothermal, self-gravitating,
ideal MHD turbulence with different rms Mach numbers $\Ms$ and physical
sizes $L$, chosen in such a way as to keep the ratio $\Ms/J$
constant. This implies that the mean density and rms Mach number of the
simulations satisfy Larson's (1981) scaling relations with their
physical size, so that the smaller ones can be considered to be a part
of the larger ones.  The simulations were performed with a resolution of
$512^{3}$ zones, using a total variation diminishing scheme
\citep{Kim+99} with periodic boundary conditions. The initial conditions
in all simulations have uniform density and magnetic field strength.
The turbulence is driven in Fourier space with a spectrum 
\beq
P(k) = k^6 \exp \left[-\frac{8 k}{\kpk} \right],
\label{eq:driving}
\eeq
where $\kpk = 2(2 \mathrm{\pi}/L)$ is the energy-injection wavenumber. The
driving in the main simulations is purely rotational (or
`solenoidal'), and a prescribed rate of energy injection is applied in
order to approximately maintain the rms Mach number near a nominal
value, which characterizes each run (Fig.\ \ref{fig:vrms_vs_t}).

The restriction that the simulations satisfy Larson's (1981) relations
means that each simulation's sonic Mach number must scale as $\Ms
\propto L^{1/2}$ and its mean density as $\langle n \rangle \propto
L^{-1}$. Specifically, we choose box sizes $L=$ 1, 4, and 9 pc, and
corresponding densities $n=$ 2000, 500, and 222$\pcc$. We choose a
temperature $T = 11.4$ K, implying an isothermal sound speed of $\cs =
0.27 \kms$, and we set the turbulence driving as to produce rms Mach
numbers $\Ms=$ 5, 10, and 15, respectively. The runs are respectively
labelled Ms5J2, Ms10J4, and Ms15J6.

Table \ref{tab:run_parameters} summarizes these parameters, together
with the Jeans number $J$, and other physical quantities characterizing
our runs, such as their mass $M$, Jeans length $\LJ$, free-fall time
$\tff$, the time at which gravity is turned on, $\tgrav$, and their {\it
rms} velocity dispersion, $\vrms$. Note that, because the density--size
relation implies a constant column density, all of our simulations have
the same initial, uniform column density, of $N = 6.2 \times 10^{21}
\psc$.

In addition to these runs, three other runs were performed in order
to test for convergence and for the effect of compressible, rather than
solenoidal, driving. These all have a nominal rms Mach number $\Ms = 15$
and Jeans number $J=6$, so that they correspond to run Ms15J6. These runs
had a 100\% compressible forcing, and were performed at resolutions of
$128^3$, $256^3$, and $512^3$. They are also indicated in
Table \ref{tab:run_parameters}, with mnemonic names.

An important parameter of the simulations is the so-called {\it virial
parameter}, $\alpha$, defined as the ratio of twice the kinetic energy
to the magnitude of the gravitational energy \citep{BM92} and which, for
a spherical geometry, reads:
\begin{equation}
\alpha \equiv \frac{2 \Ek}{|\Eg|} \approx \frac{M \sigma^2}{G M^2/L} = 
\frac{3}{4\mathrm{\pi}^2} \frac{\Ms^2}{J^2}.
\label{eq:alpha_M/J}
\end{equation}
The last equality gives $\alpha$ in terms of the nondimensional
parameters $J$ and $\Ms$. For the nominal values of these parameters for
our simulations, we see that all three of our simulations have a nominal
value of $\alpha \approx 0.475$.\footnote{FK12 propose to use a
cell-to-cell estimator for the virial parameter instead of equation
(\ref{eq:alpha_M/J}), which is based on global flow parameters. We
discuss in Appendix \ref{app:alpha_param} our reasons not to use their
suggested prescription.}

The choice of magnetic parameters for
our runs requires some further discussion. Observational
results \citep[e.g.][]{MG88, Crutcher+10} suggest that the magnetic
field strength is roughly independent of density for densities below
$\sim 10^3 \pcc$, with magnitudes of a few tens of $\mu$G. We thus
choose a constant magnetic field strength for all three simulations,
since their mean densities are comparable to or below this
threshold. Moreover, since the simulations all have the same initial
column density, this choice implies that they all have the same
mass-to-flux ratio (recall that, for uniform conditions and cylindrical
geometry, $M/\phi = N/B$). We thus choose the same initial uniform field
strength, $B = 30.3~\mu$G for all three simulations, which implies that
they have the same normalized mass-to-flux ratio, $\mu = 1.3$.


\section{Results} \label{sec:results}

\subsection{Fraction of subsonic, super-Jeans structures}
\label{sec:subson_superJ}

As in Paper I, we measure the fraction of structures in the simulations
that are simultaneously subsonic and super-Jeans. We do this as a
function of structure size because in the adopted isothermal regime,
there is no inherent physical size scale for a given density
enhancement, and its `size' is a completely arbitrary,
observer-defined quantity (for example, through a density-threshold
criterion). Of course, larger structures will in general be more massive
and, because they in general have density profiles that resemble
Bonnor--Ebert spheres \citep{BKV03, Gomez+07}, they will eventually
appear more massive than the Jeans mass associated with their mean
density. On the other hand, larger structures
will tend to have larger velocity dispersions, as dictated by the fact
that the turbulent kinetic energy spectrum in general decays with
increasing wavenumber (i.e., the kinetic energy content decreases with
decreasing size scale). Thus, sufficiently small structures should
appear subsonic in general. The question is then whether, on average,
there is a range of scales within a turbulent supersonic flow where the
structures appear both subsonic and super-Jeans.

We consider two types of regions in the simulations: either cubic
subboxes (or `cells') of fixed sizes that fill up the numerical box,
or dense clumps defined by a density threshold criterion. The subboxes
are independent of the local density structure, and constitute just a
subdivision of the numerical box, thus providing us with a very large
statistical sample in the case of sizes significantly smaller than the
numerical box. Also, they can have average densities
larger or smaller than the mean density of the numerical simulation. We
consider subboxes of sizes 2, 4, 8, 16, 32, 64, and 128 grid zones per
side. 

On the other hand, the clumps are
exclusively overdense regions, and their shapes are dictated by the
local density structure. Thus, the sample of these structures contains
much fewer elements than the subbox sample. As in Paper I, we then
define the clumps' size as the cubic root of their volume $V$, assuming
they are spherical, so that $L = (3V/4 \mathrm{\pi})^{1/3}$, and produce a
logarithmic histogram with size bins of the form $[2^n,2^{n+1}]\Delta x$,
for $n=1, \ldots$,6, where $\Delta x$ is the grid cell size. Finally, also
following our procedure in Paper I, we identify both kinds of structures
at two different times in each run: just before the gravity is turned
on, at which the density distribution is only due to turbulent effects,
and at a time around two {\it global} free-fall times after gravity is
turned on, at which the density structure is influenced by both
turbulence and gravity. Note that, when we speak of `super-Jeans'
structures at the times when self-gravity has not been turned on yet, we
simply mean that their masses are larger than the corresponding Jeans
mass at the structure's mean density and temperature.

\begin{figure*}
\includegraphics[scale=1.]{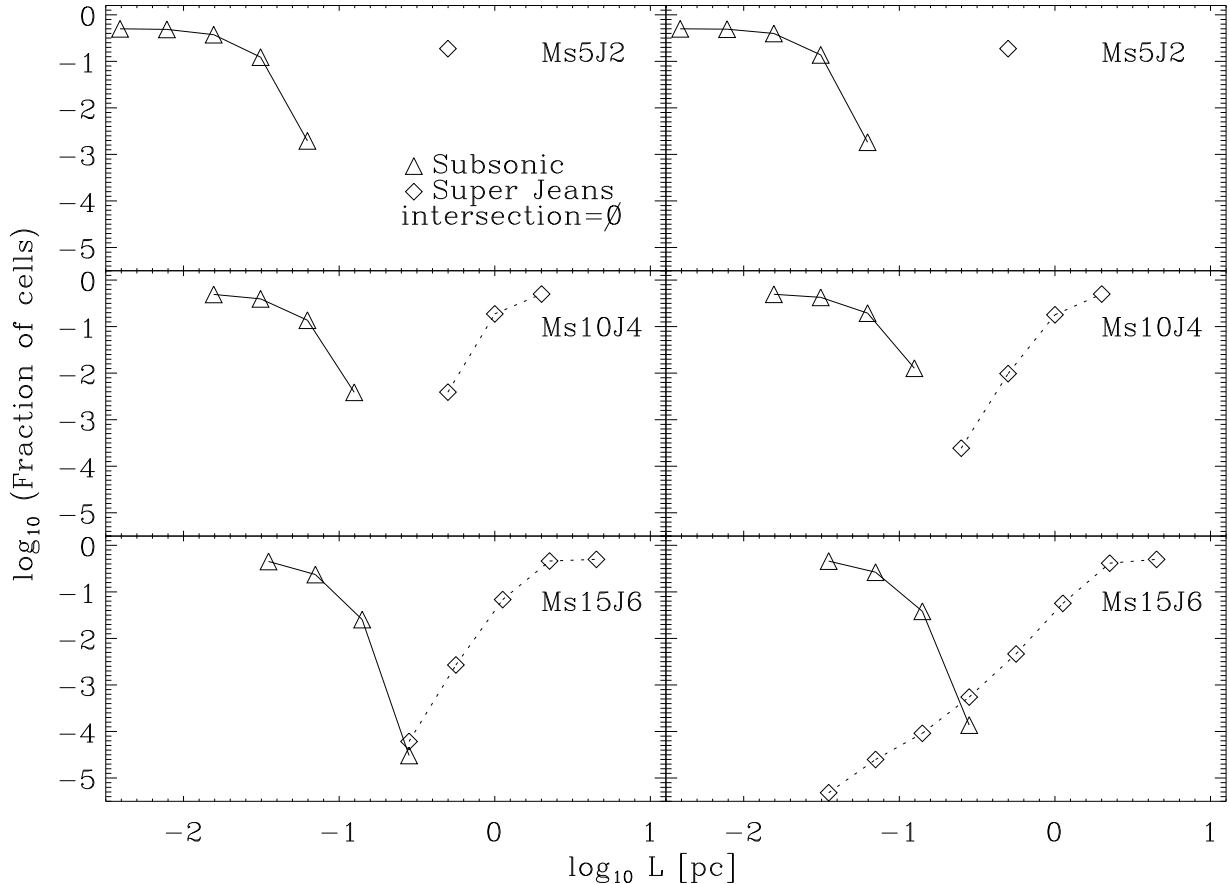}
\caption{Fraction of subsonic (triangles, solid lines) and super-Jeans
(diamonds, dotted lines) subboxes (in logarithmic scale) for runs Ms5J2
(top panels), Ms10J4 (middle panels), and Ms15J6 (bottom panels), as a
function of subbox size. The left-hand panels show the fractions
shortly before the time $\tgrav$ when self-gravity is turned on. The
right-hand panels show the fractions at approximately one free-fall time
after $\tgrav$. The fraction of subboxes that are both subsonic and
super-Jeans is zero at all subbox sizes, and thus cannot be shown in
this figure.}
\label{cells}
\end{figure*}

\begin{figure*}
\includegraphics[scale=1.]{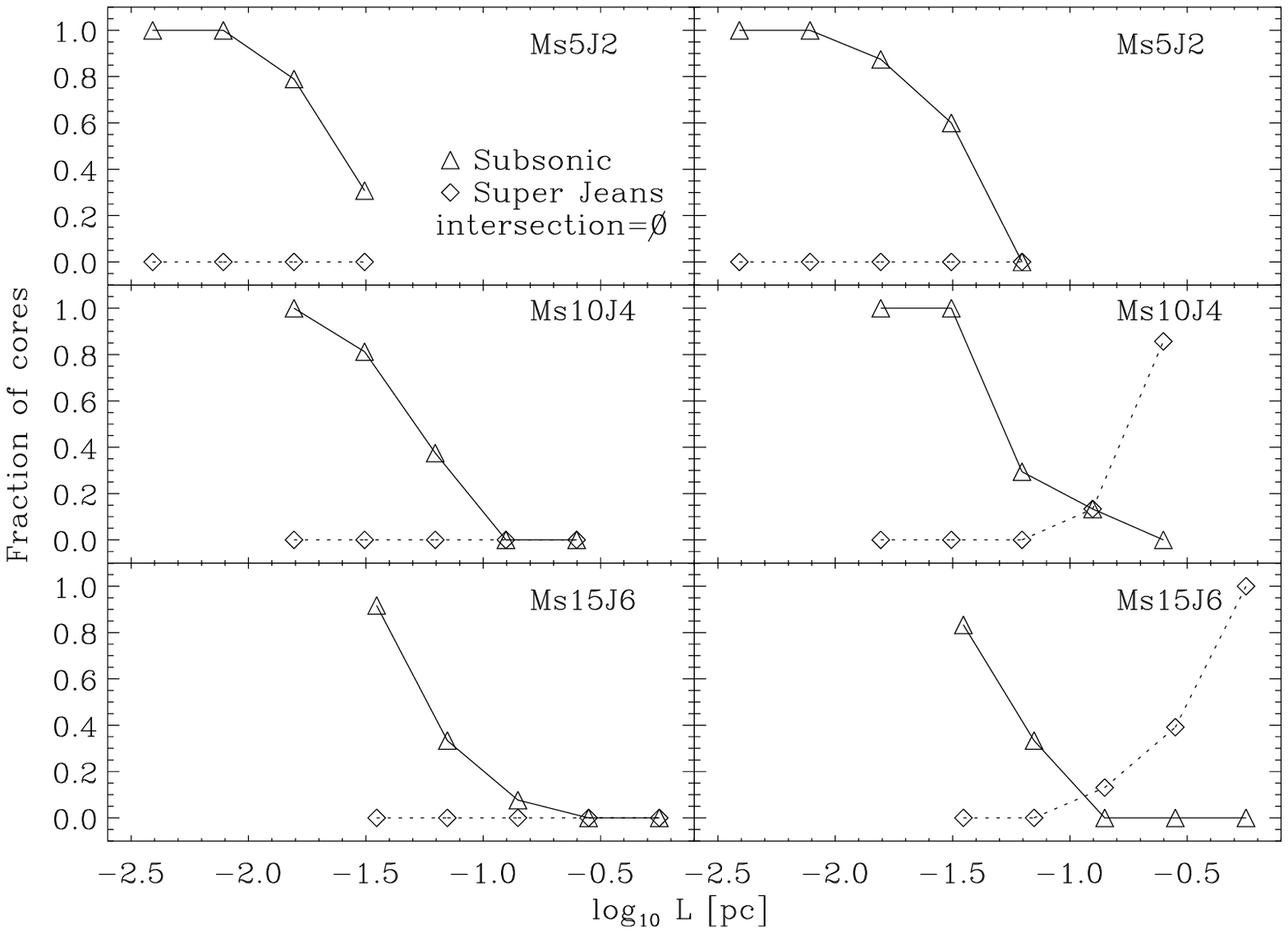}
\caption{Fraction of simultaneously subsonic (triangle, solid lines)
and super-Jeans (diamonds, dotted lines) clumps in runs Ms5J2, Ms10J4,
and Ms15J6, as a function of clump sizes. The clumps are defined as
connected regions above a certain density threshold. The ensemble of
clumps was created by considering thresholds 32, 64, 128, and 256 times
the mean density $n_0$. The fraction of clumps that are both subsonic
and super-Jeans is zero at all clump sizes considered.}
\label{cores}
\end{figure*}

Figs \ref{cells} and \ref{cores}, respectively, show the fractions of
subsonic ({\it solid lines} and {\it triangles}) and of super-Jeans structures
({\it dotted lines} and {\it diamonds}) for subboxes and clumps, as a
function of their sizes. In both figures, the top panels present the
results from run Ms5J2, the middle panels from run Ms10J4, and bottom
panels from run Ms15J6. The results computed before (resp.\ after)
self-gravity is turned on are shown in the left-hand (resp.\ right-hand)
panels. In Fig. 1, the fraction of subboxes is shown in logarithmic
scale because the total number of subboxes is very large for the
smallest sizes, and thus the subsonic and super-Jeans fractions can be
very small.

It is important to note that, similarly to the
case of the non-magnetic simulations presented in Paper I, we have found no
structures (either cells or cores) that are {\it simultaneously}
subsonic and super-Jeans at any of the size scales we considered in this
work. In Figs \ref{cells} and \ref{cores}, when both the subsonic and
the super-Jeans curves show non-zero values at a given scale, these
correspond to {\it different} structures, that are {\it either} subsonic or
super-Jeans, but no structure among the ones we sampled has both
properties simultaneously. 

In addition, no significant effect is observed in the fraction of
subsonic structures at a given scale upon the inclusion of self-gravity.
On the other hand, the fraction of super-Jeans structures as a function
of scale exhibits a clear change after self-gravity is turned on.
However, the effect is different for the subboxes and the clumps. For
the former, the range of scales at which super-Jeans cells exist is
stretched towards {\it small} scales when self-gravity is on. Instead,
for clumps, {\it no super-Jeans structures exist in the absence of
self-gravity}, and when it is included, super-Jeans clumps appear at
{\it large} scales.

We conclude from this section that, similarly to the non-magnetic case
studied in Paper I, simultaneously subsonic and super-Jeans structures
are uncommon also in driven, MHD supersonic, isothermal turbulent
flows. We discuss some implications of these results in Sec.\
\ref{sec:discussion}.

%
%

%
%

\subsection{Velocity convergence} \label{sec:vel_conv}

A second nearly universal assumption about the non-thermal motions in
molecular clouds and their substructure is that they consist of random
turbulence, which provides an isotropic, `turbulent' pressure, in a
similar manner to thermal motions and pressure. In Paper I, we
tested this hypothesis by computing the mean divergence in subboxes of
size equal to the physical size of the small-scale simulation within
the largest scale simulation, and plotting it against the mean density
of the regions. In that paper, we found that there exists a negative
correlation between the mean divergence of the flow and the mean density
of the subboxes, suggesting that, on average, overdense regions are
characterized by a net convergence of the velocity field within
them. Such a velocity field structure has a reduced capability of
countering the self-gravity of the structures, and in fact may be {\it
caused} by it. We now test whether this result persists in the MHD case.

As in Paper I, we subdivide the large-scale simulation, Ms15J6, into
cells of size equal to that of run Ms5J2,
and compute the mean density and mean divergence of the velocity field
for each cell. The divergence is computed by dotting the Fourier
transform of the velocity field with the corresponding wavevector
{\bf k}.

\begin{figure*}
\includegraphics[scale=0.32]{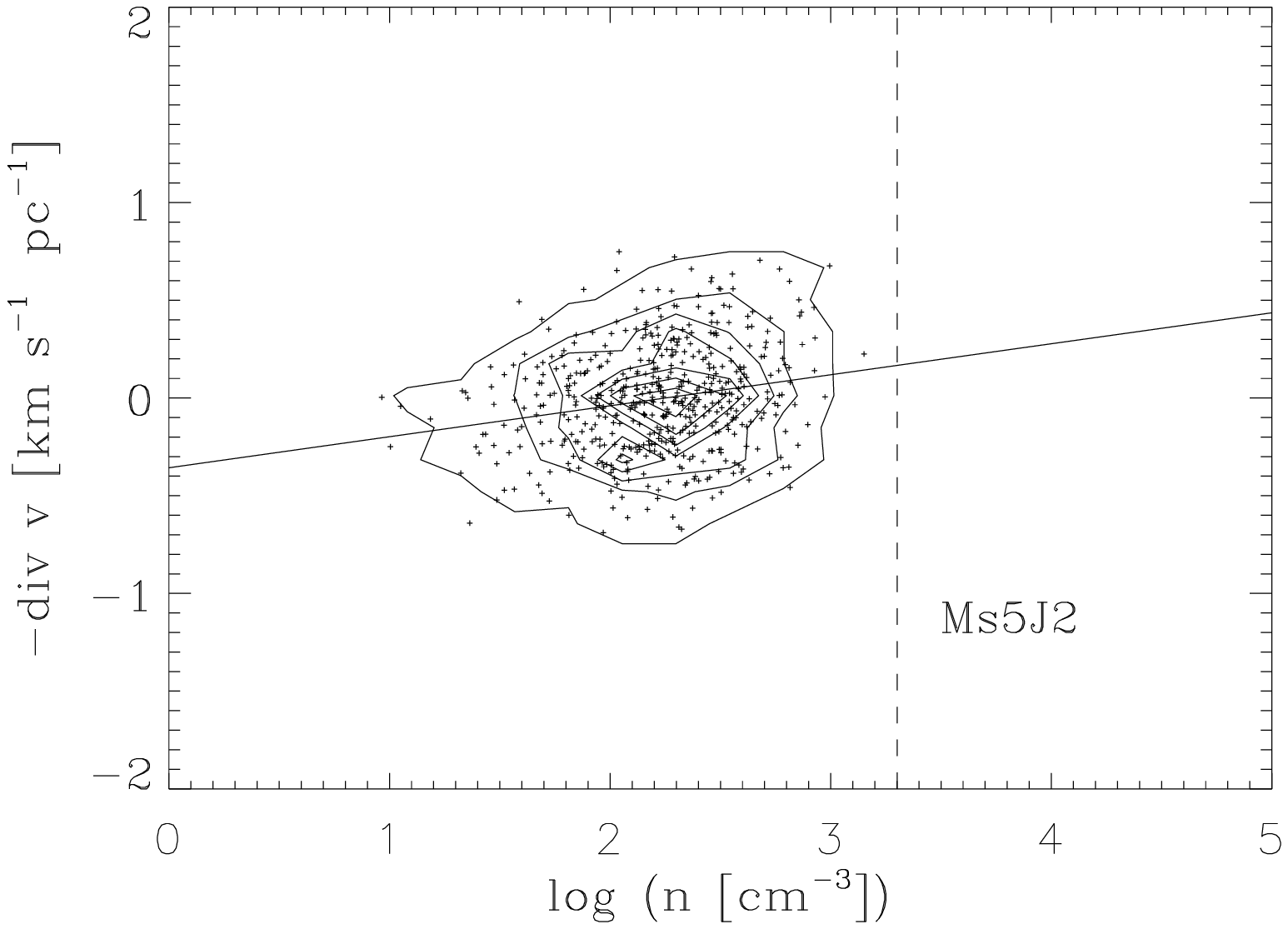}
\includegraphics[scale=0.32]{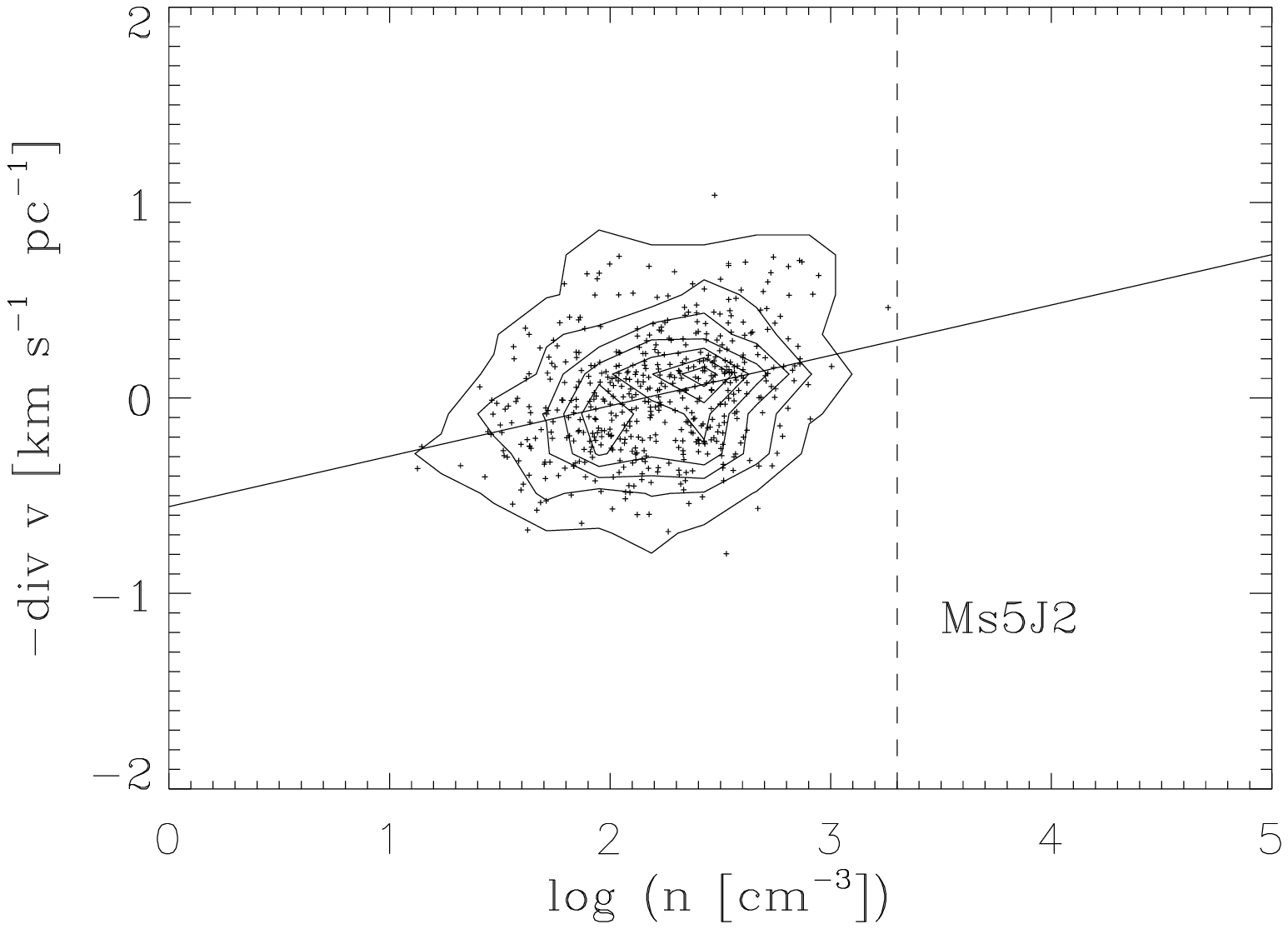}
\includegraphics[scale=0.32]{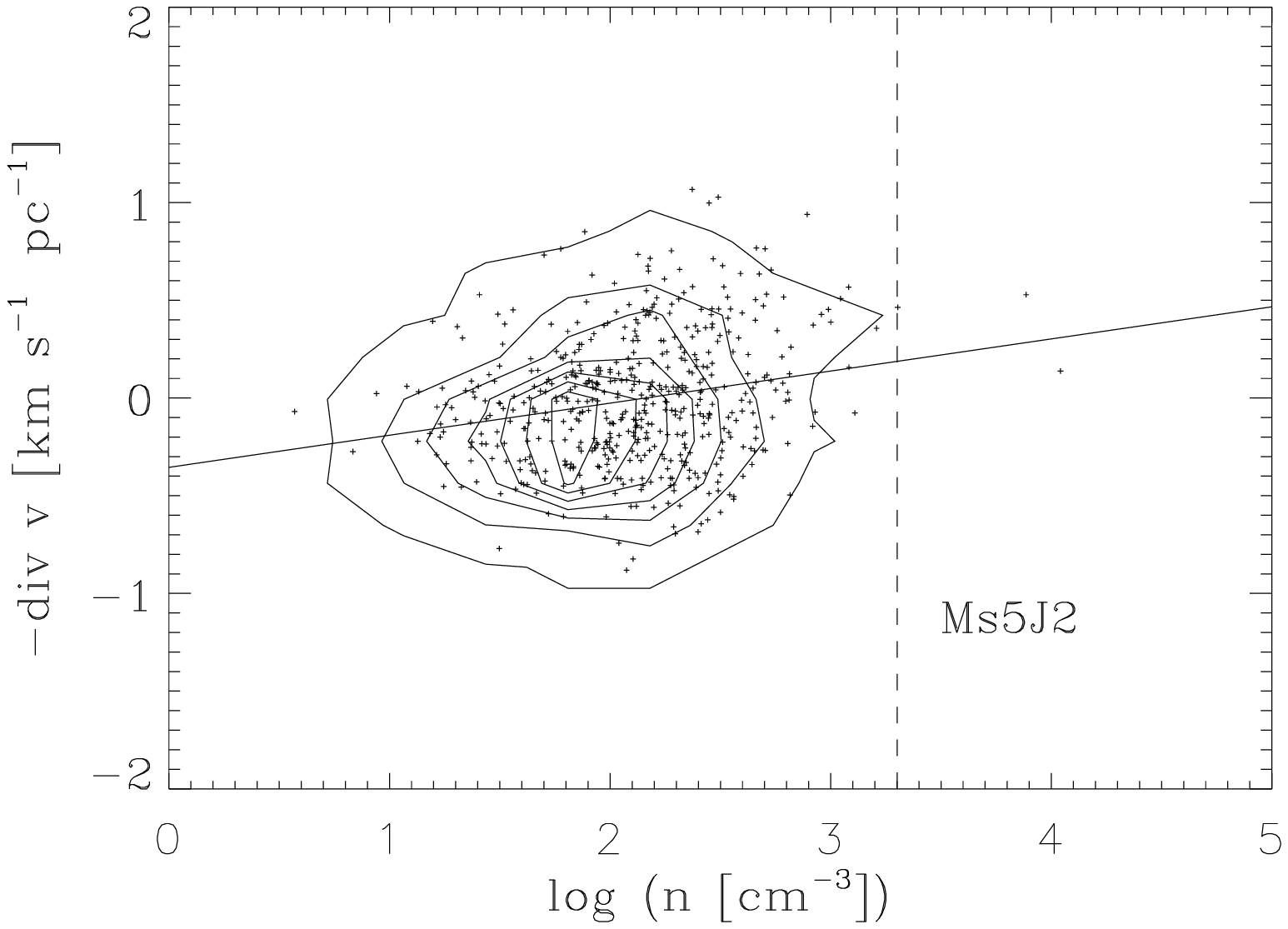}
\caption{Negative mean velocity divergence {\it versus} mean
density for subboxes of the large-scale simulation Ms15J6. The left and
middle panels show the two-dimensional histograms of the subboxes in
this plane at two times before self-gravity is turned on, namely $t = 3$
Myr, corresponding to 1 $\tturb$ (left-hand panel) and $t= 6$ Myr (2
$\tturb$; middle panel). The right-hand panel shows the histogram at $t =
13.5$ Myr, corresponding to one free-fall time after gravity was turned
on. The straight solid lines show least-squares fits through the data
points, and have slopes 0.16 (left-hand panel), 0.26 (middle panel), and 0.16
(right-hand panel), with respective correlation coefficients of 0.21, 0.31,
and 0.22.  The contours are drawn at increments of 1/7th of the
maximum. The dashed vertical lines show the mean density of the small-scale 
simulation Ms5J2.}
\label{div1}
\end{figure*}


Fig.\ \ref{div1} shows the two-dimensional histogram of the cells in
the $\{-\nabla \cdot {\bf v}$, $\mbox{log}\, n \}$ space, at three different
times during the simulation. The left and middle panels, respectively,
show the histogram at $t= 3$ and 6 Myr,
which, respectively, correspond to 1 and 2 $t_{\rm turb}$, where $\tturb
\equiv L/v_{\rm rms} =$ 3 Myr is the turbulent crossing time (see
Table \ref{tab:run_parameters}). Both of these correspond to times
before self-gravity was turned on. The right-hand panel shows the histogram
at $t= 13.5$ Myr, which corresponds to one free-fall time after gravity
was turned on. The contours are drawn at levels 1/14, 1/7, 2/7,$\ldots $, 6/7 of the maximum. We find in all cases that the contours show an
elongated shape, and the straight solid lines show least-squares fits
through the data points. The fitted slope has values of $\sim 0.16\pm
0.033$ at time $t=3$ Myr, $\sim 0.26 \pm 0.035$ at $t=6$ Myr, and $\sim 0.16
\pm 0.33$ at $t=13.5$ Myr, with corresponding correlation coefficients
0.21, 0.31, and 0.22. The uncertainties reported are the 1$\sigma$
errors of the fit.  However, given the low correlation coefficients, and
the fact that the earliest and latest distributions exhibit
approximately the same slope, the slopes can be considered to be the
same before and after turning self-gravity on, $\sim 0.19$, with the
main difference being that the distribution of points in this diagram
becomes more elongated in the presence of self-gravity, as evident
from the larger extent to higher densities of the lowest contour in this
case (right-hand panel of Fig.\ \ref{div1}).

\begin{figure*}
\includegraphics[scale=0.45]{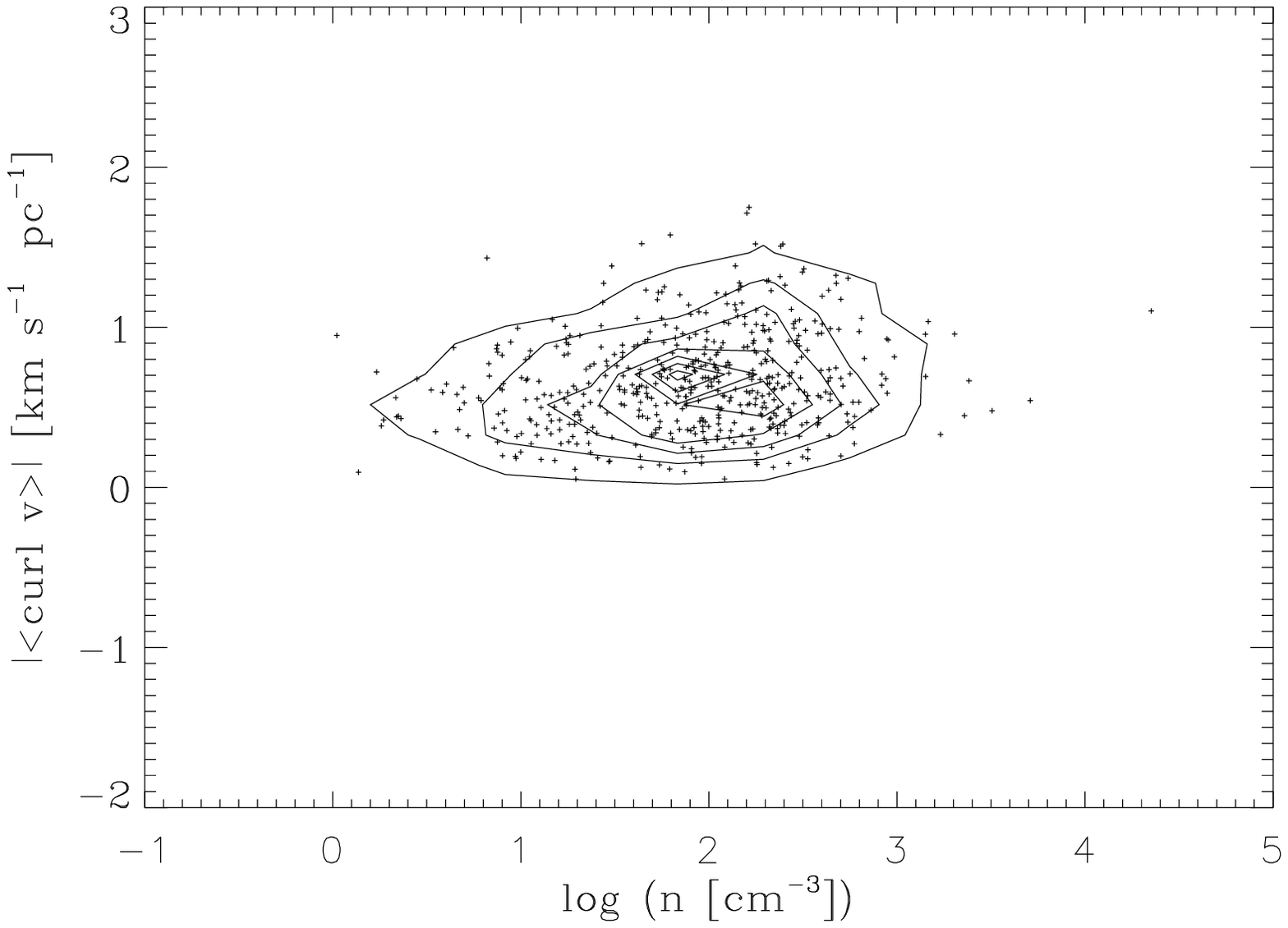}
\includegraphics[scale=0.45]{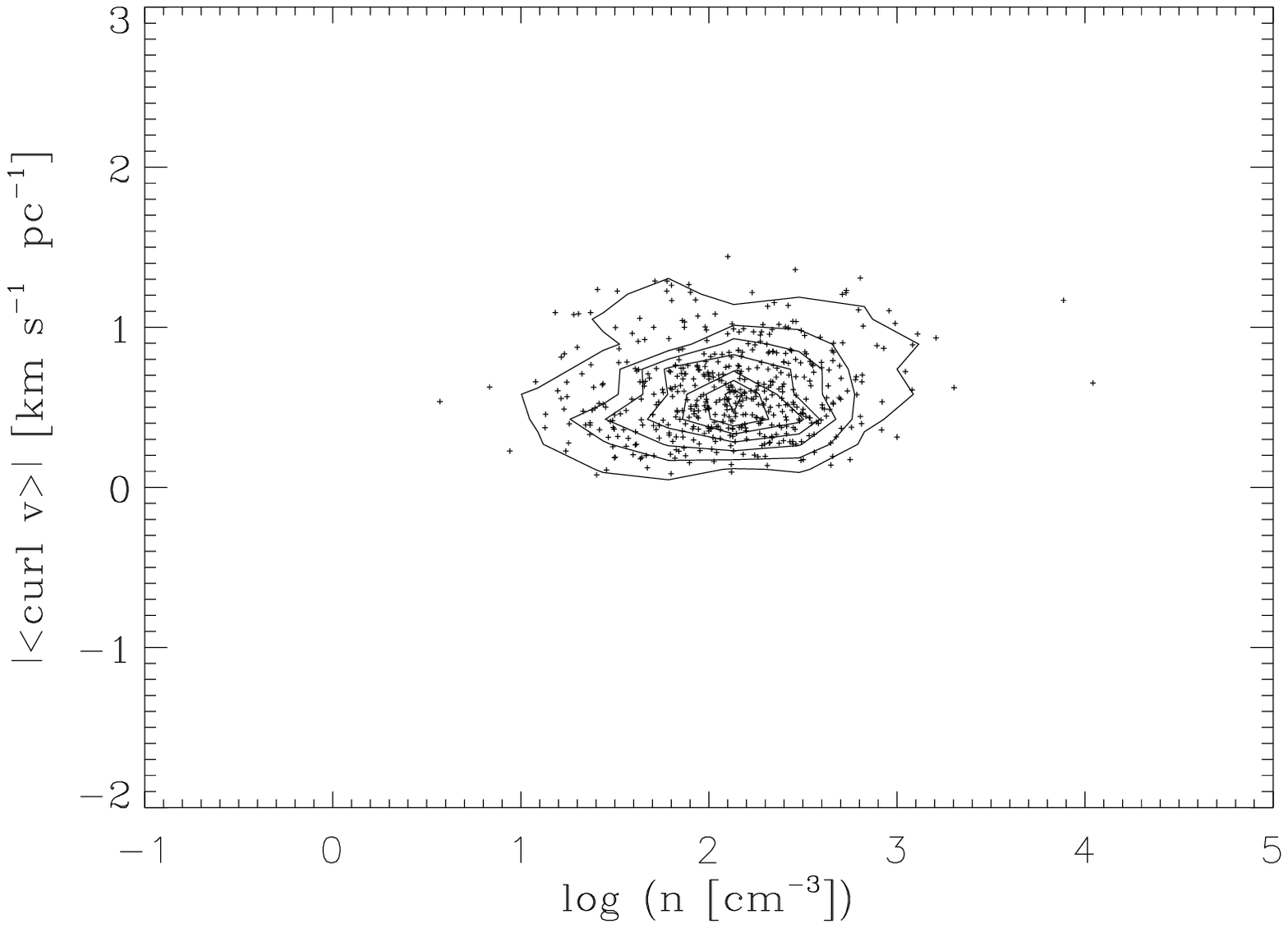}
\caption{Magnitude of the velocity curl {\it versus} mean
density for subboxes of the large-scale simulation Ms15J6 in the
non-magnetic case (left-hand panel; from Paper I) and in the magnetic case
(right-hand panel; this paper). The range of $\nabla \times \vv$ values is
seen to be very similar in both cases, while the density range is seen
to be significantly shorter in the magnetic case.}
\label{fig:curl_vs_rho}
\end{figure*}

Compared to our result from Paper I in the non-magnetic case, in
which we had found slopes ranging from 0.35 to 0.5, 
the present MHD simulations exhibit a weaker dependence of $\nabla \cdot
\vv$ on $n$. One possible explanation for this would be that the
magnetic field converts a larger fraction of the converging motions into
vortical ones than do the pure hydrodynamical non-linear interactions
\citep{VS+96, VS+98, RG12}.  To test this possibility, in Fig.\
\ref{fig:curl_vs_rho} we show the two-dimensional histograms in the $\nabla
\times \vv$ -- $\langle n \rangle$ space for the subboxes of run
Ms15J6 in both the non-magnetic case (left-hand panel, from Paper I) and in
the magnetic case studied in this paper (right-hand panel). Contrary to the
aforementioned expectation, it is seen that the range of $\nabla \times
\vv$ values is in fact similar in the magnetic and non-magnetic
cases. This suggests that the decrease in the slope of the $\nabla \cdot
\vv$ -- $\langle n \rangle$ correlation is {\it not} due to a
more efficient transformation of compressive motions into rotational
motions inside the dense regions in the presence of the magnetic field,
but rather, simply to a stronger average opposition to the turbulent
compressions due to the added pressure from the magnetic field
\citep[e.g.,] [] {Molina+12}.

In any case, a net correlation is still
observed between density and velocity convergence, supporting the result
from Paper I that density enhancements in a turbulent flow must on
average contain a net convergent component of the velocity field which,
rather than opposing gravitational contraction, contributes to it, or is
driven by it, even in the presence of a magnetic field of magnitude
typical of molecular clouds.
It is important to mention that we do not find any significant change in the distributions showed in Fig.\ \ref{div1} with the action of self-gravity, as can be seen by comparing the left and middle panels of Fig.\ \ref{div1} with the right one. In this figure, we plotted a line showing the density of simulation Ms5J2 and, similar to that in Paper I, we find that self-gravity appears to be necessary for the production of regions dense enough as to fall on the same Larson density--size relation (same intercept) as their parent structure, while turbulence alone seems essentially incapable of doing it.

\subsection{SF efficiency per free-fall time in
constant-virial parameter structures} \label{sec:SFE}

As in Paper I, we now use our numerical simulations to assess the
dependence of the `star formation efficiency per free-fall time',
$\sfeff$,\footnote{Note that KM05 called this quantity the `star
formation {\it rate} per free-fall time', and this nomenclature has
become common in the literature. However, this is actually somewhat
misleading, since the $\sfeff$ is obtained by integrating the SFR over
a free-fall time and then dividing by the total mass, giving the
fraction of the gas mass converted into stars over one free-fall time,
thus being an efficiency, not a rate. At best, it can be considered an
{\it average} SFR over a free-fall time, with mass normalized to the
system mass. Thus, we prefer to call it the star formation efficiency
per free-fall time.} on the turbulent rms Mach number.

To measure $\sfeff$ in our simulations, we compute the evolution of the
total mass in collapsed regions. Operationally, we define a collapsed
region as a connected region with density above a threshold density
$n_{\rm thr} = 1000~n_0$, where $n_0$ is the mean density, given in
Table \ref{tab:run_parameters} for each one of the runs. As explained in
Paper I, this is a sufficiently high density that it cannot be reached
by turbulent compressions alone, and thus it has to be the result of a
local gravitational collapse event.\footnote{Note that sink particles
are not implemented in the version of the code used in this
paper. Therefore, a gravitational collapse event simply leads to the
accumulation of all the mass involved in it in a few grid cells. Also,
this implies that the `collapsed object', defined by $n_{\rm thr}$,
does not have a strictly constant mass, because it can oscillate around its
equilibrium state, and thus have a variable mass above $n_{\rm
thr}$. Resolution is not a concern here, because we are only concerned
with the total collapse mass, and not about how it is distributed into
fragments, which would be the affected outcome of the collapse in the
case of insufficient resolution \citep{Truelove+97}.} 
It was shown in Paper I that using a threshold $n_{\rm thr} = 500~n_0$
does not significantly alter the results.

\begin{figure*}
\includegraphics[scale=0.49]{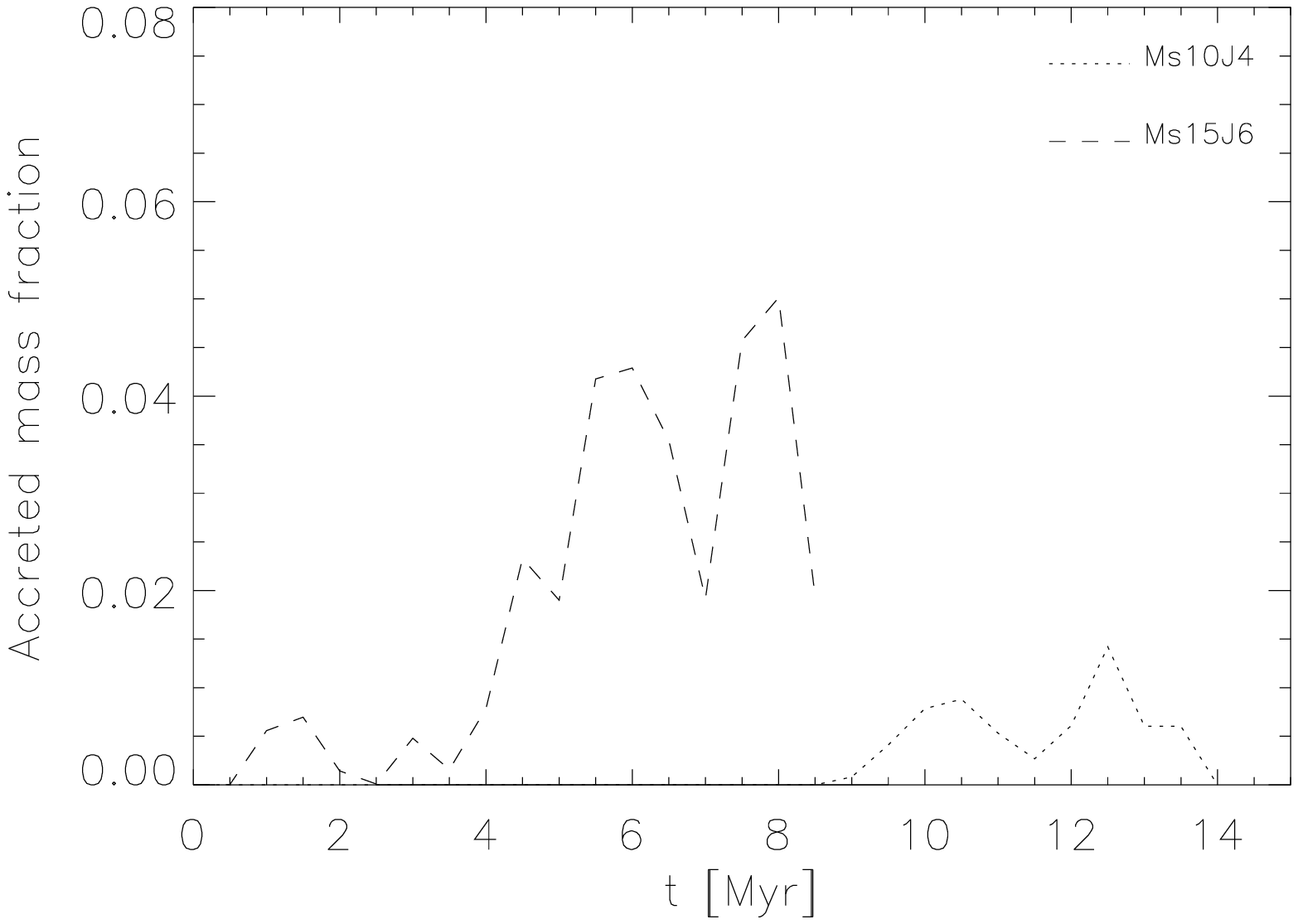}
\includegraphics[scale=0.49]{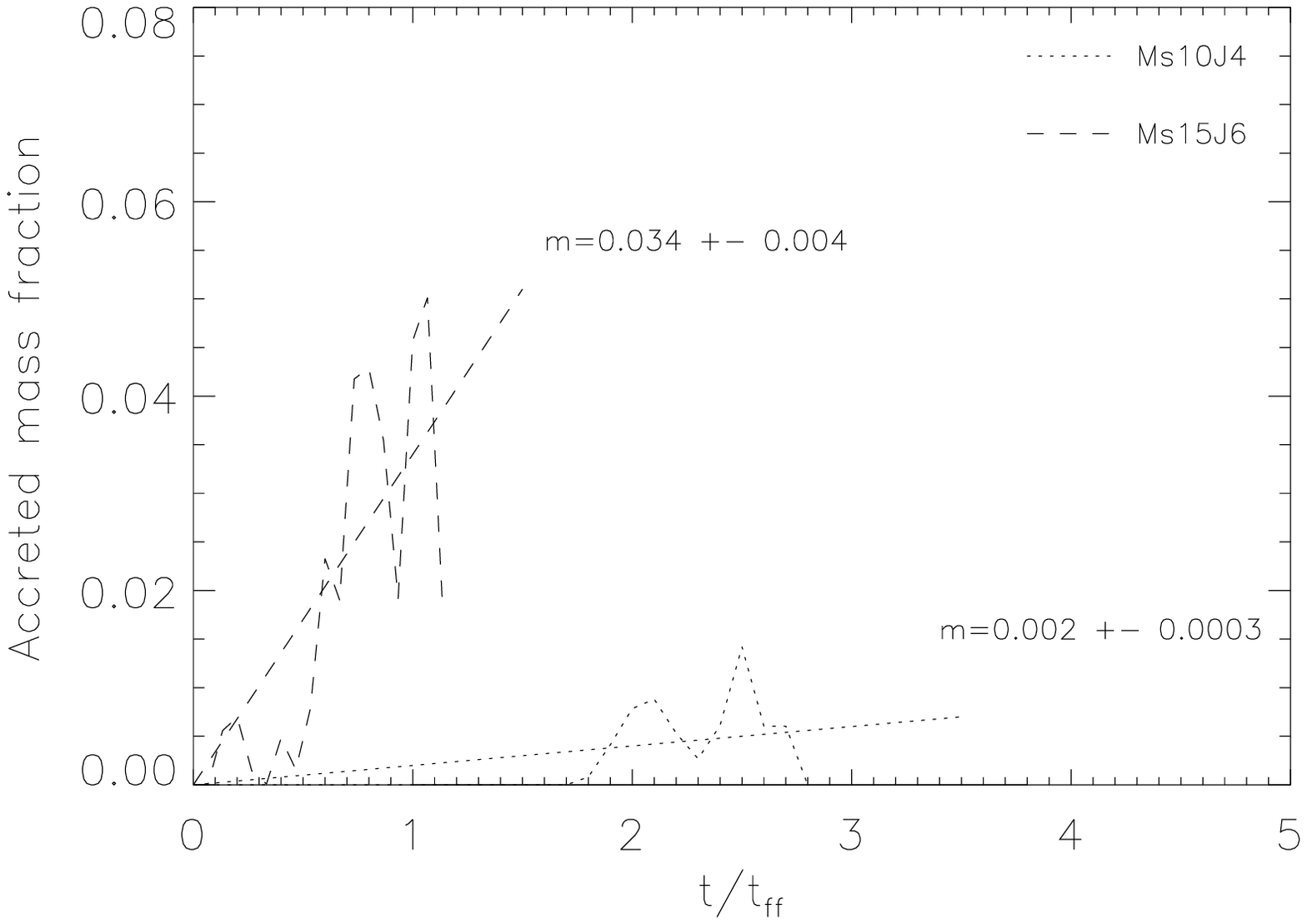}
\caption{Fraction of mass accreted into collapsed objects as function of time
in units of Myr (left-hand panel) and of the free-fall time $\tff$ (right-hand panel) 
for the simulations Ms10J4 and Ms15J6. Run Ms5J2 is not shown in this
figure because there is no accreted mass in this simulation. The average
$\sfeff$ is computed by fitting a least-squares line to the plot of accreted
mass versus time in units of $\tff$.}
\label{amvst}
\end{figure*}

\begin{figure*}
\includegraphics[scale=0.65]{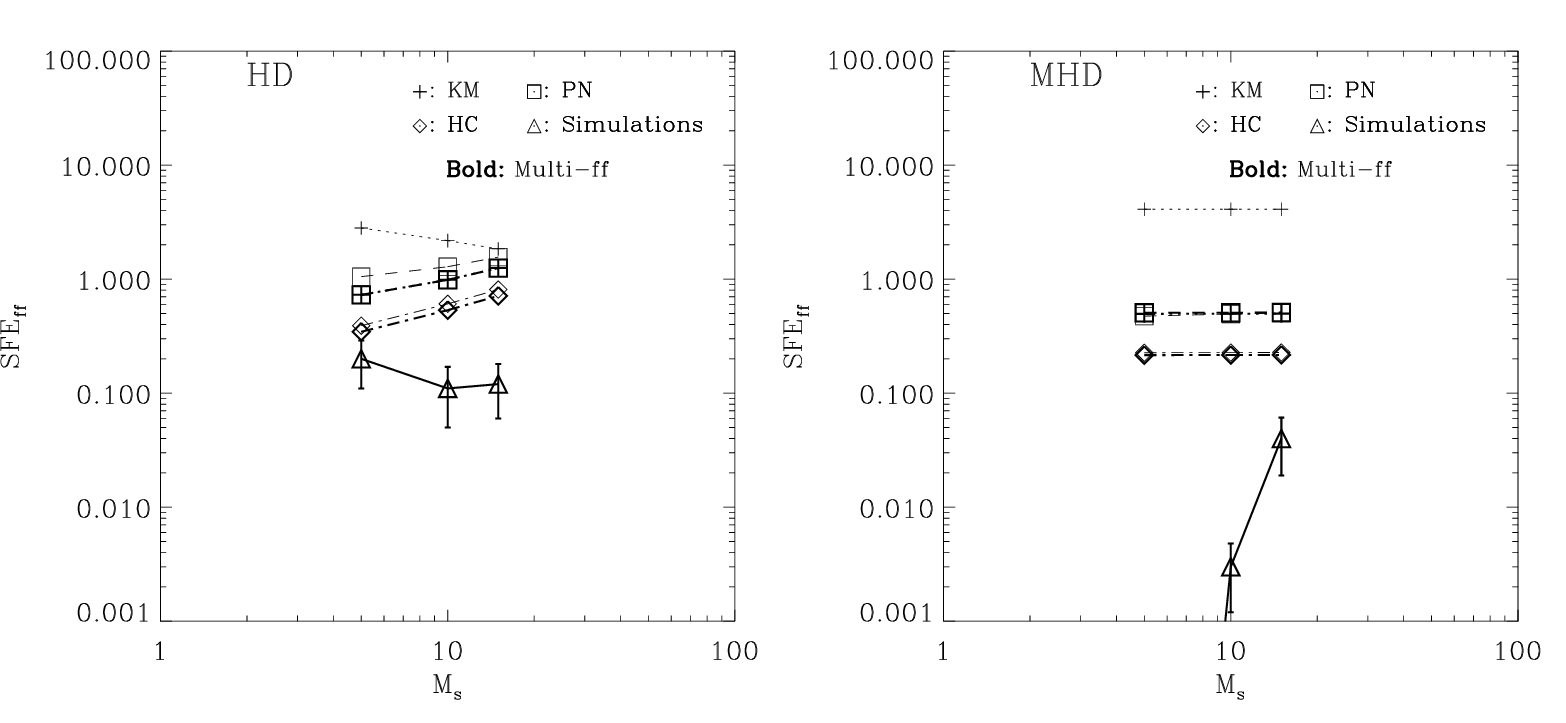}
\includegraphics[scale=0.65]{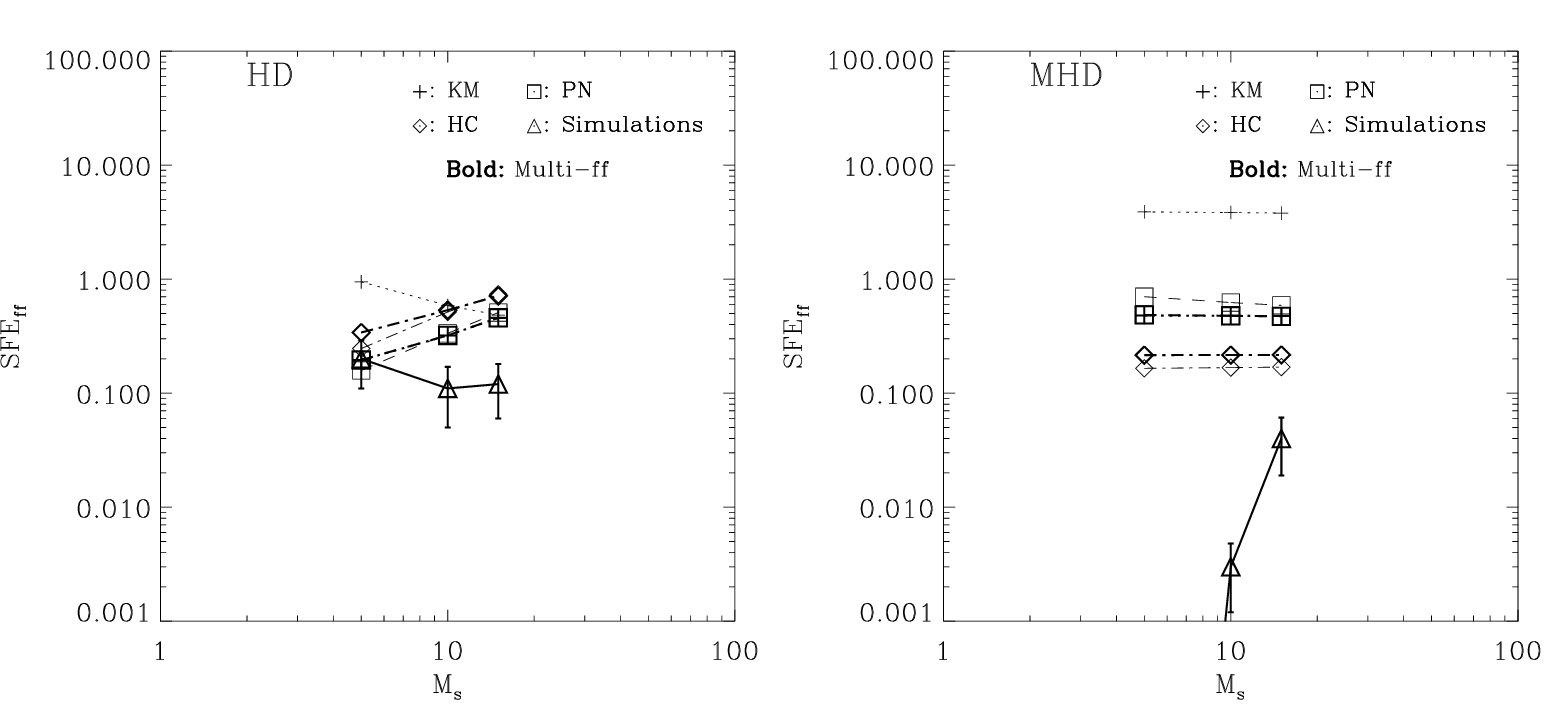}
\caption{{\it Top-left panels:} $\sfeff$ {\it versus} rms Mach number for the
three non-magnetic simulations from Paper I (which used the same values
of the parameters as the ones presented here, except for the absence of
the magnetic field), and for the predictions from the theories by KM05,
PN11, and HC11, in both their original as well as their `multi-free-fall' modes (shown in bold lines and symbols), according to the
expressions given by FK12 in the non-magnetic limit ($\beta \rightarrow
\infty$). The error bars in the values for the simulation data indicate
$3 \sigma$ errors. {\it Top-right panels:} same as the left-hand panel, but
showing the results from the magnetic simulations Ms10J4 and Ms15J6, as
well as the predictions from the six models, according to the extension
to the magnetic case proposed by FK12. Run Ms5J2 in the magnetic case
produced no collapsed objects, and is indicated by the line going off
the plot down and to the left from the point for run Ms10J4. {\it Bottom
panels:} same as the respective top panels, but showing the model
predictions assuming a value of the $\alpha$ parameter 10 times larger
than the nominal one, to consider the possible effect of measuring it
directly from the simulation, as done by FK12.}
\label{fig:sfeff}
\end{figure*}

Fig.\ \ref{amvst} shows the accreted mass fraction as function of
time for runs Ms10J4 and Ms15J6, starting from the time at which gravity
was turned on in each case. Run Ms5J2 is not shown in this figure
because, contrary to the non-magnetic case, no gravitational collapse
occurs in this simulation. When the time is written in units of the
simulation free-fall time, the slope of this curve gives $\sfeff$. In
the right-hand panel of Fig.\ \ref{amvst}, the dashed and dotted lines show
least-squares fits to the evolution of the accreted mass for runs Ms15J6
and M10J4, respectively, with their slopes indicated. We observe that
$\sfeff = 0.002 \pm 0.0003$ for run Ms10J4 and $\sfeff = 0.034 \pm 0.004$
for run Ms15J6, where the indicated uncertainties are the 1-$\sigma$
errors of the linear fit, due to the noisiness of the accreted mass
graphs and the scarcity of collapsed objects. These values are shown, as
a function of the corresponding rms Mach number of the simulations, by
the solid line and triangles in the {\it right-hand panel} of Fig.\
\ref{fig:sfeff}. Note that $\sfeff = 0$ for run Ms5J2, and thus this run
is off the plot.

As discussed in Sec.\ \ref{sec:num_sim} and in Appendix
\ref{app:alpha_param}, FK12 have suggested that the virial parameter be
measured directly from the simulation data, rather than from the global
run parameters. Upon doing that, they obtained values of $\alpha$ up to
an order of magnitude larger than the nominal, global ones. As discussed
in the appendix, it is unclear to us how realistic these values are, and
thus which one is a better choice but, as simple test, we show in the
bottom panels of Fig.\ \ref{fig:sfeff} the model predictions assuming
$\alpha = 4.75$, i.e., a value 10 times larger than the nominal one,
to obtain a feel for the $\sfeff$ predicted by the models in this
case. It is seen that, for the non-magnetic case, the larger value of
$\alpha_{\rm vir}$ implies somewhat smaller values of the $\sfeff$,
ameliorating the discrepancy between the predicted and measured values
of the $\sfeff$, although still the only model that captures the {\it
trend} of $\sfeff$ {\it versus} $\Ms$ (the multi-free-fall KM05 model)
differs by nearly an order of magnitude from the simulation
measurement. In the case of the PN11 and HC11 models, although
agreeing in absolute value with the measurement of the $\sfeff$ at low
$\Ms$, exhibit the opposite trend with $\Ms$, and so they are off the
measured value by nearly an order of magnitude at the highest value of
$\Ms$. On the other hand, in the magnetic case, the change in $\alpha$
introduces almost no variation in the model predictions for the $\sfeff$
in the magnetic case, and the discrepancy with the simulation
measurements remains.

\subsection{Absence of collapse in run Ms5J2} \label{sec:run_Ms5J2}

As already mentioned above, no collapse occurred in run Ms5J2, in spite
of it being both Jeans unstable, with a Jeans length equal to half the
simulation box size, and magnetically supercritical, with a
mass-to-flux ratio 1.3 times critical (cf.\ Table
\ref{tab:run_parameters}). This is illustrated in Fig.\ \ref{rhomax},
which shows the evolution of the maximum density in the three runs (top
panels), and also for their non-magnetic counterparts from Paper I. It
is seen that run Ms5J2 has not produced collapsed regions even after
three and a half free-fall times, and that run Ms10J4 takes nearly two
free-fall times to develop a collapse, while run Ms15J6 produces it
immediately after self-gravity is turned on.

In order to understand this behaviour, we consider the combined effect of
the two supporting agents, thermal pressure and magnetic field, in a
virial balance calculation. As is well known, the virial equilibrium for
a uniform-density spherical mass of radius $R$ in hydrostatic
equilibrium, supported by thermal pressure and the magnetic field, is
\citep[see, e.g.,][]{Shu92}
\beq
3 \int P dV + \frac{1}{8 \pi} \int B^2 dV - \frac{3}{5}
\frac{GM^2}{R} = 0,
\label{eq:vir_eq}
\eeq
which leads to an equilibrium, `magneto-thermal Jeans
radius'\footnote{FK12 also use this quantity, although they omit the
factor of 3 in the thermal contribution, because they added in
quadrature the Alfv\'en velocity and the sound speed to obtain total
thermal+magnetic pressure in the medium. Instead, we obtain the extra
factor of 3 because we consider the virial balance of the cloud.}
\beq
\Rjeff = \left[\frac{5\cs^2(3+1/\beta)}{4 \pi G \rho}\right]^{1/2},
\label{eq:eff_Jeans}
\eeq
where we have used the isothermal equation of state $P=\cs^2 \rho$, the
definition of the Alfv\'en speed, equation (\ref{b0}), and the fact that
$\beta = 2 (\cs/\va)^2$. We then can see that the magneto-thermal Jeans
radius is a factor $f = \left[(3+1/\beta)/3\right]^{1/2}$ times larger
than the pure thermal value, which corresponds to $\beta \rightarrow
\infty$. In turn,
we can thus define the {\it magneto-thermal Jeans parameter} $\Jeff =
J/f$, which is given in the last column of Table
\ref{tab:run_parameters} for the three runs. Specifically, it is 0.91,
0.95, and 1.02 for runs Ms5J2, Ms10J4, and Ms15J6, respectively. Thus,
although the difference between the three cases is only $\sim 10$\%,
this parameter is indeed minimum for run Ms5J2, for which it appears
that the global magnetic support is enough to prevent collapse at least
over more than three simulation free-fall times. Correspondingly, at $\Jeff
\approx 0.95$, magnetic support is able to delay the occurrence of
collapse for nearly two free-fall times for run Ms10J4, while at $\Jeff
\approx 1.02$ for run Ms15J6, magnetic support seems to already make
essentially no difference with respect the non-magnetic case.

\begin{figure*}
\end{figure*}

\begin{figure*}
\includegraphics[scale=0.49]{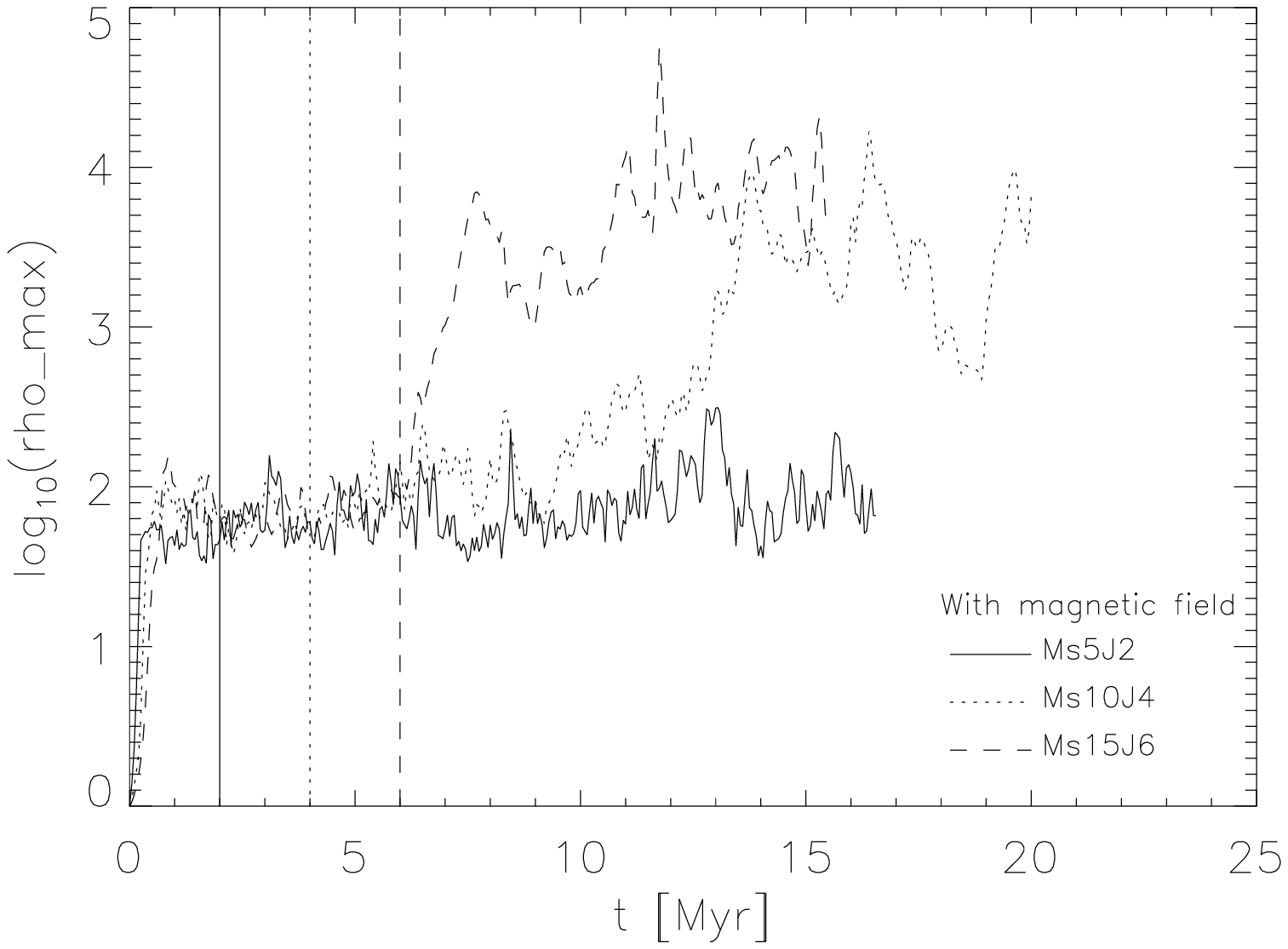}
\includegraphics[scale=0.49]{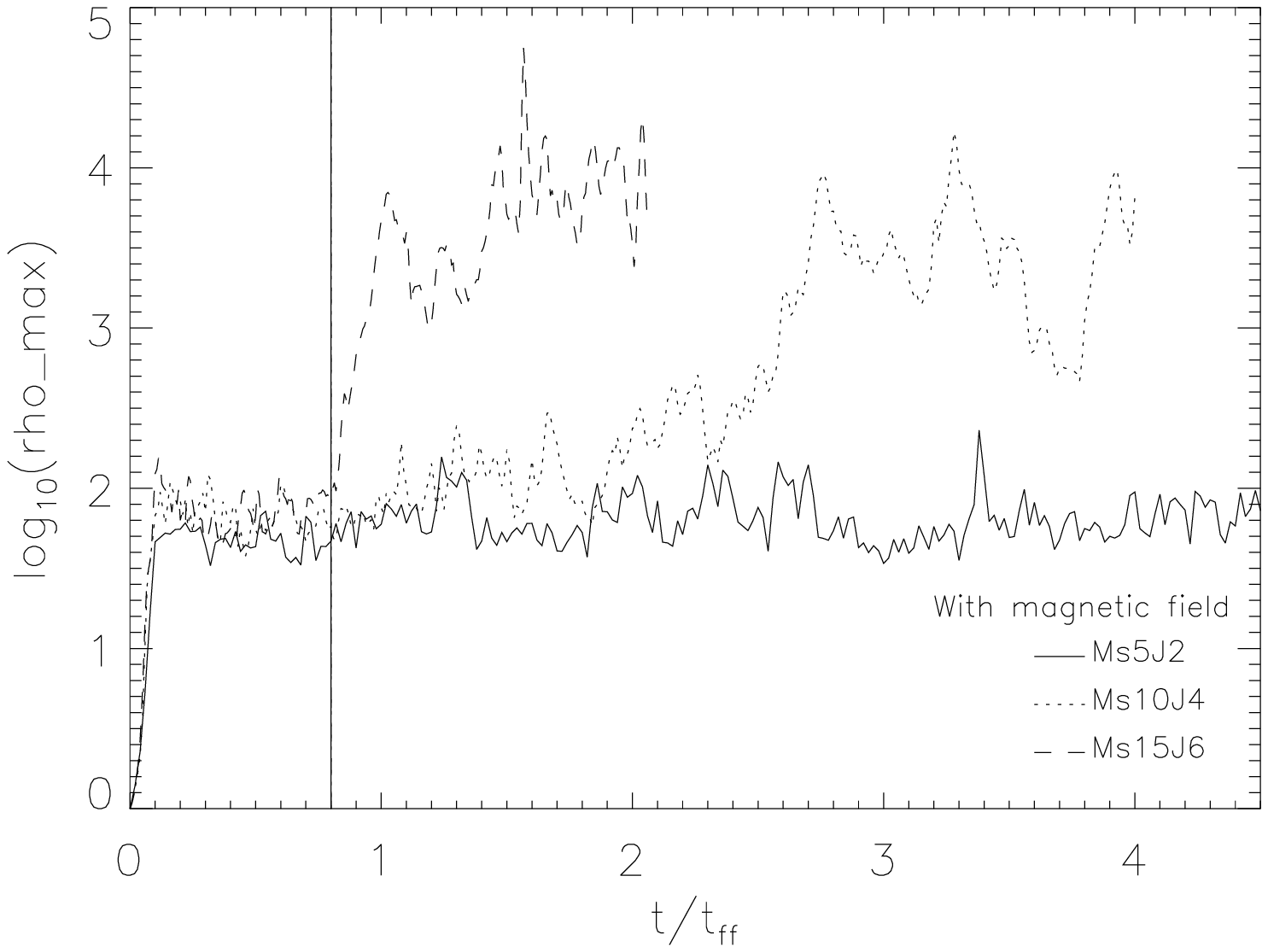}
\includegraphics[scale=0.49]{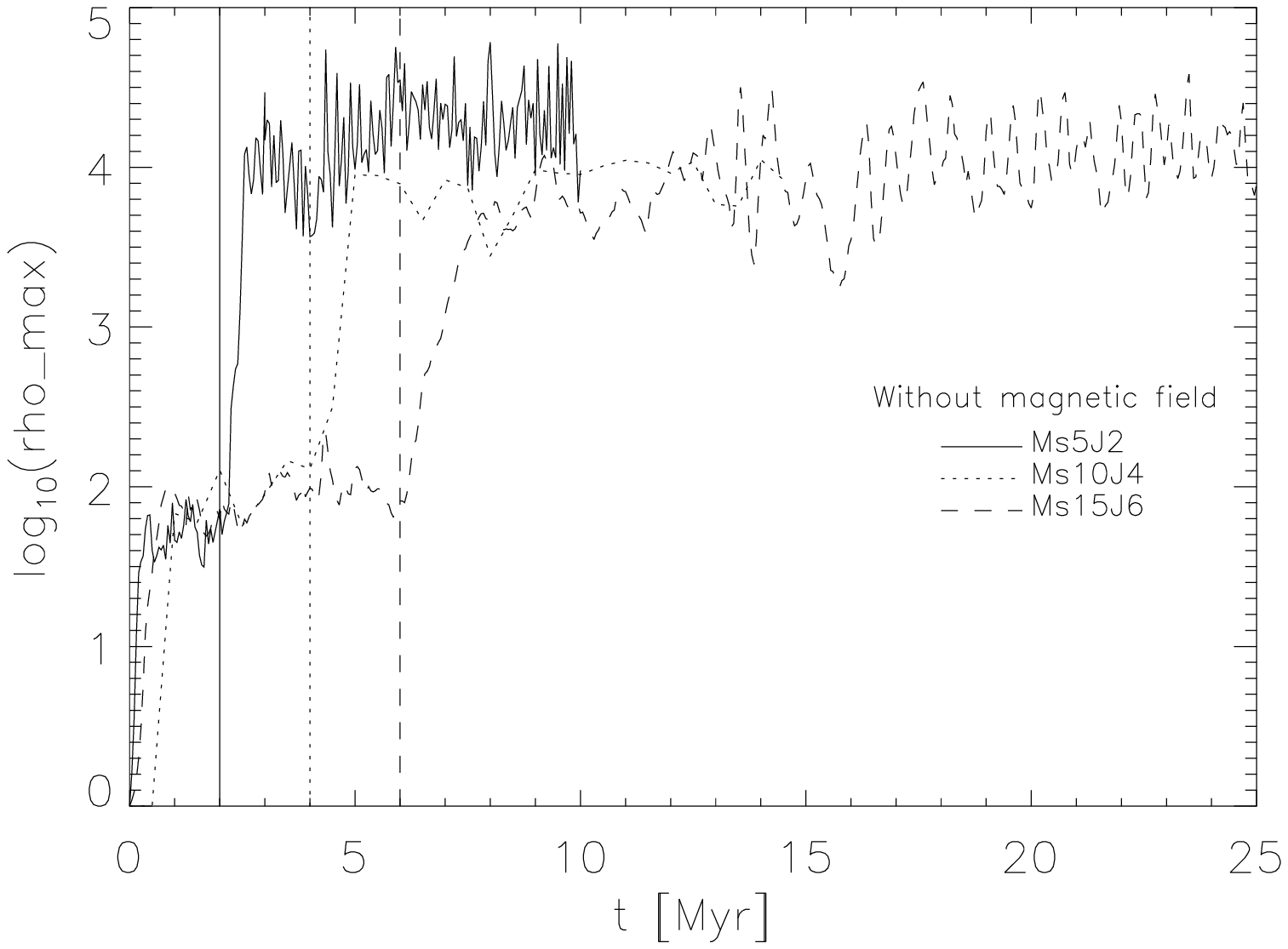}
\includegraphics[scale=0.49]{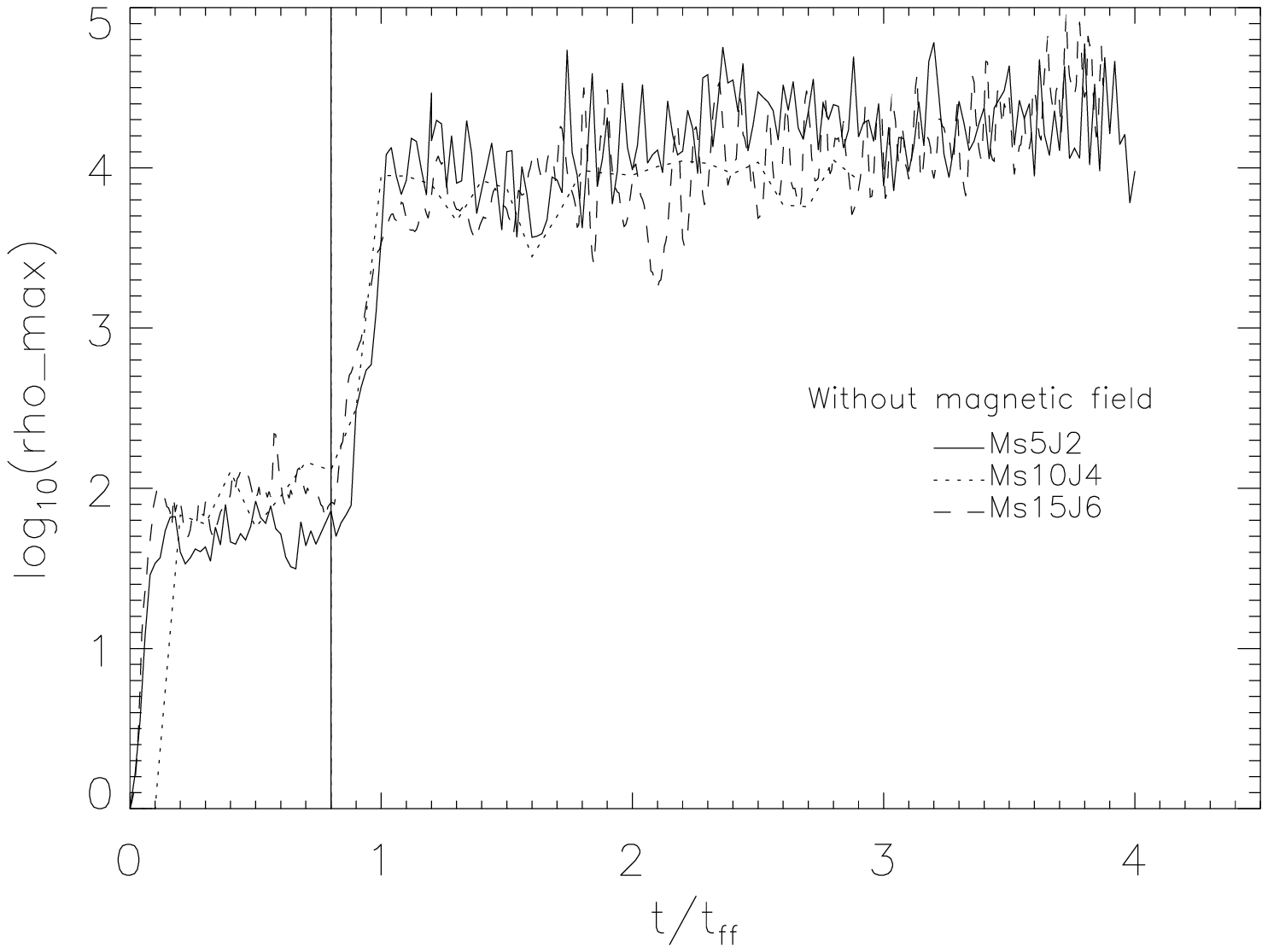}
\caption{Maximum density as function of time for the three magnetic
simulations in this paper (top panels), and for their non-magnetic
counterparts from Paper I (bottom panels). The left-hand panels show the time
in Myr, while the right-hand panels show it in terms of the simulation
free-fall time. The vertical lines show the time $t_{\rm grav}$ when
self-gravity is turned on for each simulation, which is the same
in units of the free-fall time for the three runs.}
\label{rhomax}
\end{figure*}

\subsection{Effects of resolution and type of driving}
\label{sec:resol_comp_driv}

The results discussed so far refer to simulations performed at a fixed
resolution of $512^3$ and with solenoidal driving, following the scheme
used in Paper I. It is important, however, to test whether they are
affected by the numerical resolution and whether they hold in the
presence of compressible driving. In this section, we discuss the
results from a few additional simulations designed to test this.

\begin{figure*}
\includegraphics[scale=1.]{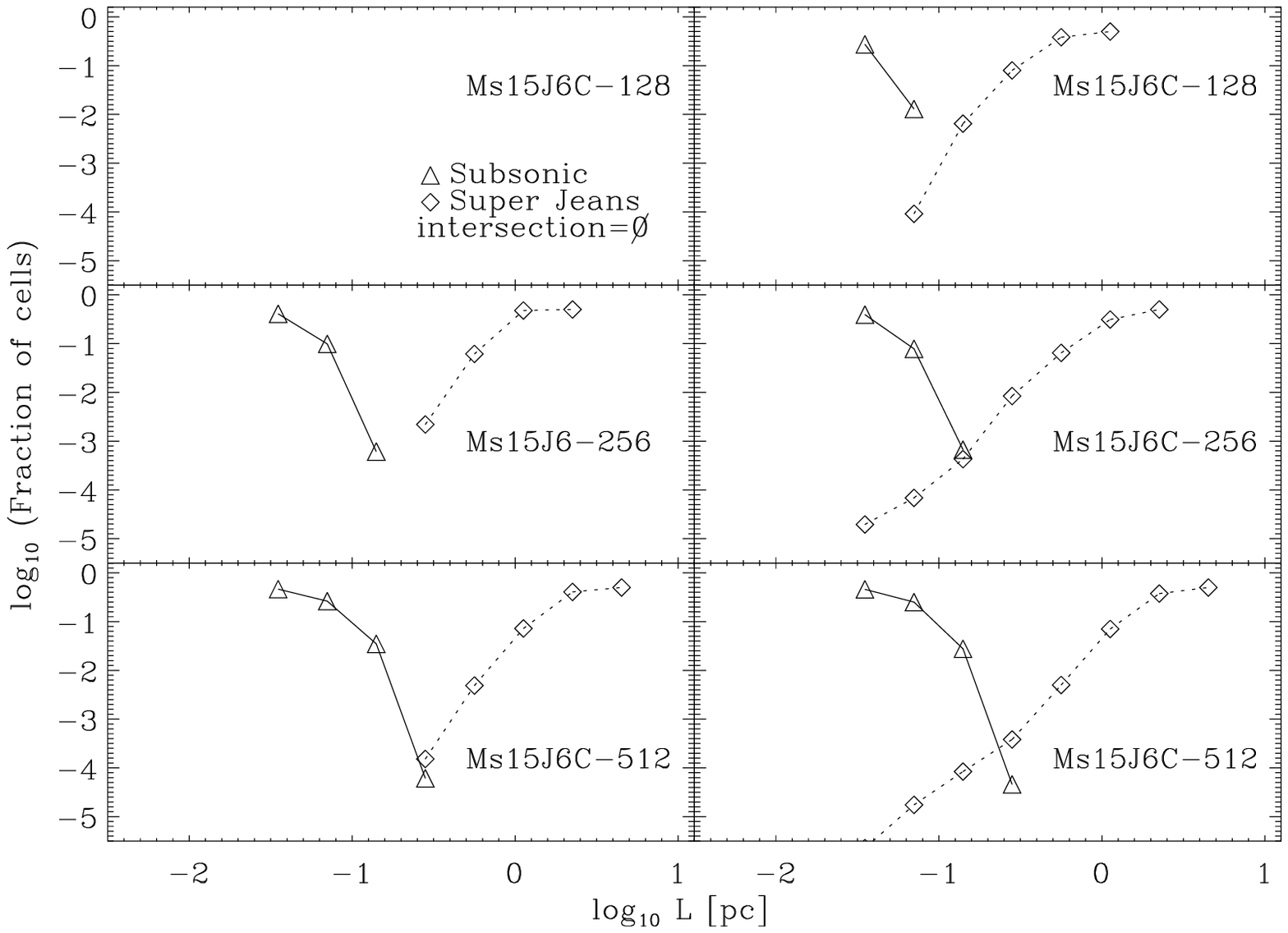}
\caption{Fraction of simultaneously subsonic (triangles, solid lines)
and super-Jeans (diamonds, dotted lines) subboxes (in logarithmic
scale) for the compressible-driving runs Ms15J6C-128 (top panels),
Ms15J6C-256 (middle panels), and Ms15J6C-512 (bottom panels), as a
function of the subbox size. The left-hand panels show the fractions
shortly before the time $\tgrav$ when self-gravity is turned on. The
right-hand panels show the fractions at approximately one free-fall time
after $\tgrav$. The fraction of subboxes that are both subsonic and
super-Jeans is zero at all subbox sizes, and thus cannot be shown in
this figure.}
\label{fig:sprJ_subson_comp_cells}
\end{figure*}

\begin{figure*}
\includegraphics[scale=1.]{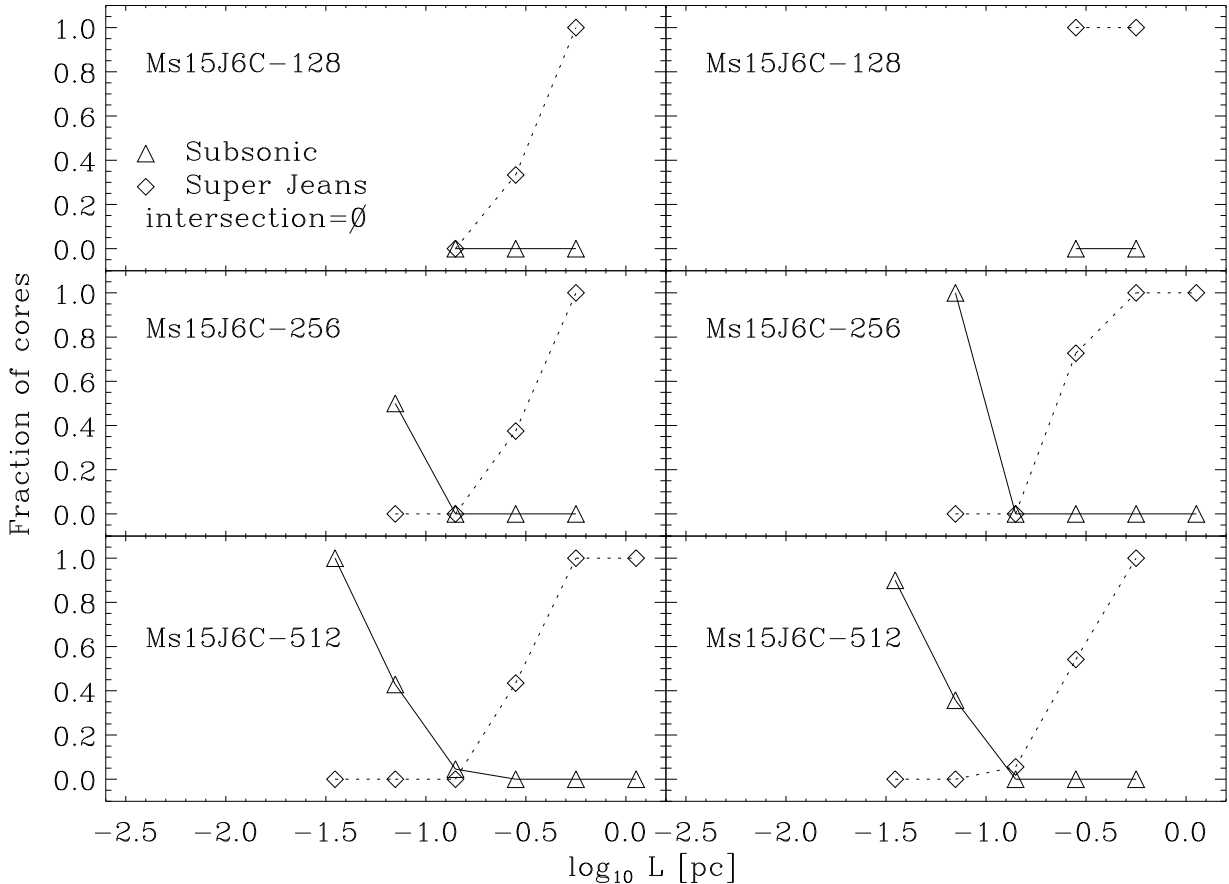}
\caption{Fraction of simultaneously subsonic (triangle, solid lines)
and super-Jeans (diamonds, dotted lines) clumps in the
compressively driven runs Ms15J6C-128, Ms15J6C-256, and Ms15J6C-512, as a
function of clump sizes. The clumps are defined as connected regions
above a certain density threshold. The ensemble of clumps was created by
considering thresholds 32, 64, 128, and 256 times the mean density
$n_0$. The fraction of clumps that are both subsonic and super-Jeans is
zero at all clump sizes considered.}
\label{fig:sprJ_subson_comp_cores}
\end{figure*}

In Fig.\ \ref{fig:sprJ_subson_comp_cells}, we show the fraction of
subsonic and of super-Jeans subboxes (cf.\ Fig.\ \ref{cells}) in the
numerical box for the compressible runs. Fig.\
\ref{fig:sprJ_subson_comp_cores} shows the corresponding result for the
cores in these simulations. It is seen that at all three resolutions,
the compressible runs also do not exhibit simultaneously subsonic and
super-Jeans structures, neither subboxes of the simulation nor clumps
selected as density enhancements above a certain density
threshold. Thus, this result from the solenoidal simulations continues
to hold when the forcing is fully compressible. Of course, we cannot
rule out that, at higher resolution, such structures may appear, but
we defer higher resolution simulations to a future study, and perhaps
using an adaptive-mesh code. So, here we can only report that, up to our
highest resolution, such structures do not appear, regardless of the
compressibility of the driving applied.

\begin{figure*}
\includegraphics[scale=1.]{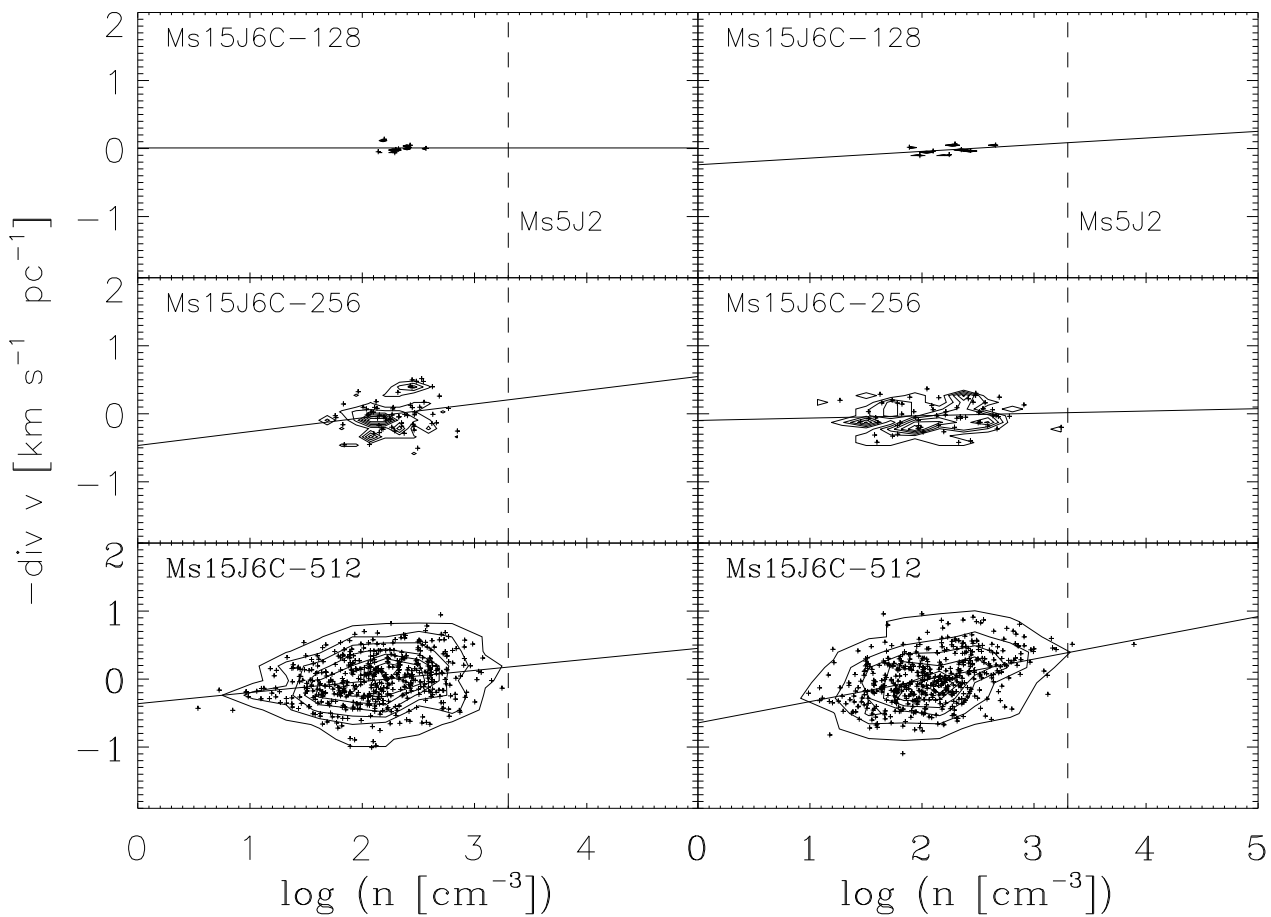}
\caption{Negative mean velocity divergence {\it versus} mean
density for subboxes of the large-scale simulation with compressible
forcing, Ms15J6C, at the three resolutions we considered. The left-hand
panels show the two-dimensional histograms of the subboxes in this
plane at time $t = 6$ Myr, before self-gravity is turned on,
corresponding to 2 $\tturb$ after the start of the simulations
(left-hand panel). The right-hand panels show the histogram at $t = 13.5$
Myr, corresponding to one free-fall time after gravity was turned
on. The contours are drawn at increments of 1/7th of the maximum. The
straight solid lines show least-squares fits through the data
points. For the high-resolution run Ms15J6C-512, the fits have slopes
$0.16 \pm 0.031$ (left-hand panel) and $0.31 \pm 0.034$ (right-hand panel), with
correlation coefficients 0.22 and 0.38, respectively. The dashed vertical lines show the mean density of the small-scale 
simulation Ms5J2.}
\label{fig:div_comp}
\end{figure*}

Fig.\ \ref{fig:div_comp} shows the two-dimensional histograms of the
cells in the $\{-\nabla \cdot \vv$, $\mbox{log}\, n \}$ space for the
three compressively driven runs, before (left-hand panels) and after (right-hand
panels) having turned self-gravity on. While the correlations are poorly
defined at the low resolutions, it can be seen that, for the
highest resolution run Ms15J6C-512, the distribution is qualitatively
very similar to that for the solenoidally driven run Ms15J6 (cf.\ Fig.\
\ref{div1}). In particular, the fitted slopes for run Ms15J6C-512 are
$0.16 \pm 0.031$ and $0.31 \pm 0.034$, with correlation coefficients
0.22 and 0.38, for the distributions before and
after turning self-gravity on, respectively, thus spanning a very
similar range to that observed in run Ms15J6. Moreover, the scatter in
the distributions is also similar, so we conclude that this result is
also independent of whether the driving is solenoidal or compressible.

\begin{figure*}
\includegraphics[scale=0.75]{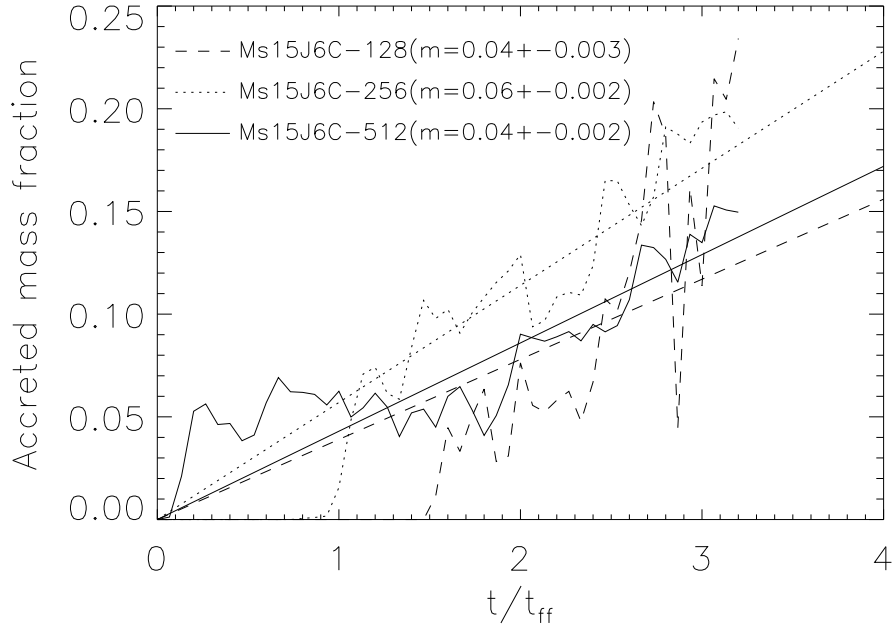}
\caption{Fraction of mass accreted into collapsed
objects as function of time in units of the free-fall time $\tff$ for
the compressive-forcing simulations Ms15J6C-128, Ms15J6C-256, and
Ms15J6C-512. The straight lines show least-squares fits to the accretion
histories, with the indicated fitted slopes. }
\label{fig:accr_hist_comp}
\end{figure*}

\begin{figure*}
\includegraphics[scale=0.55]{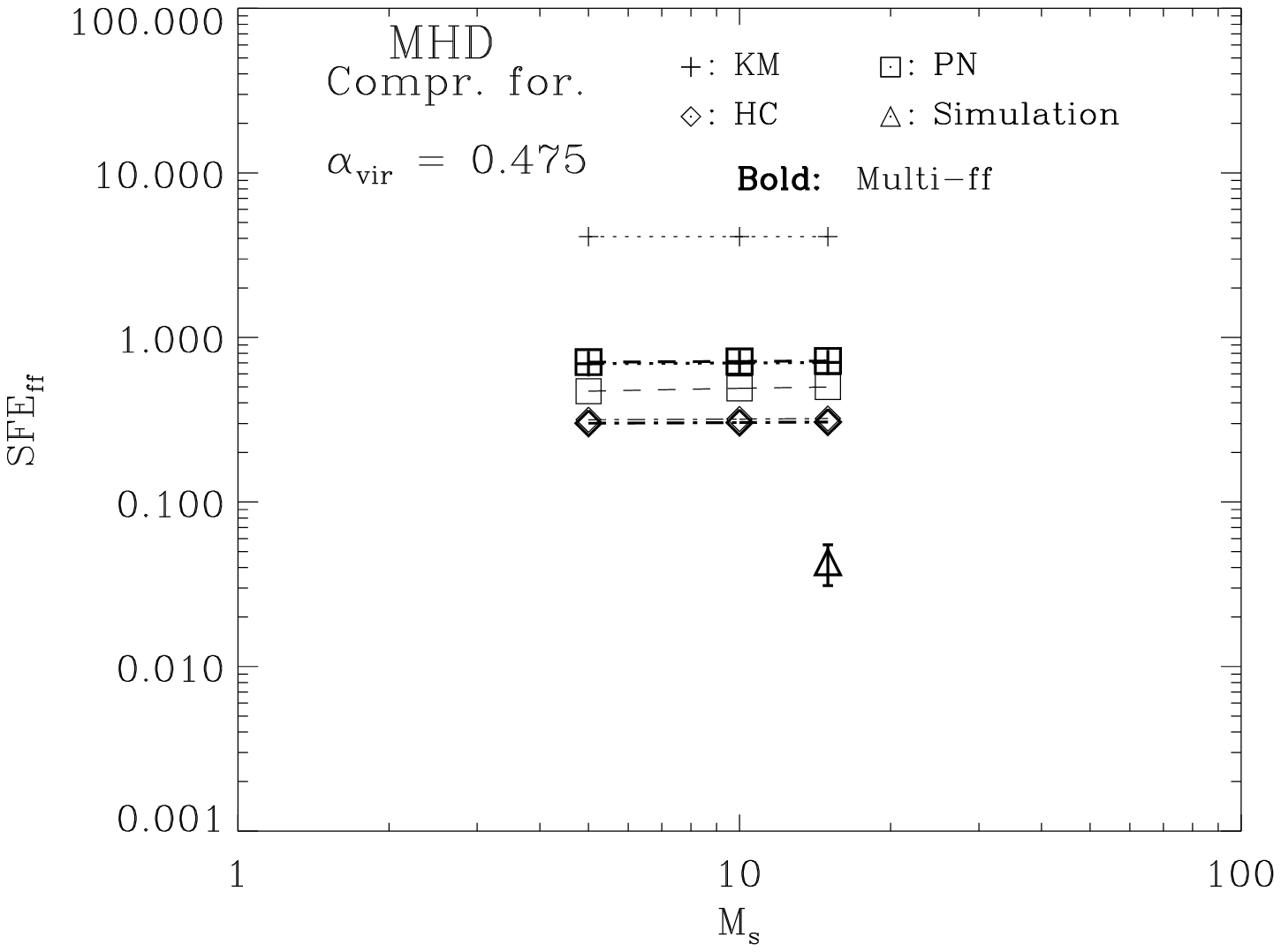}
\includegraphics[scale=0.55]{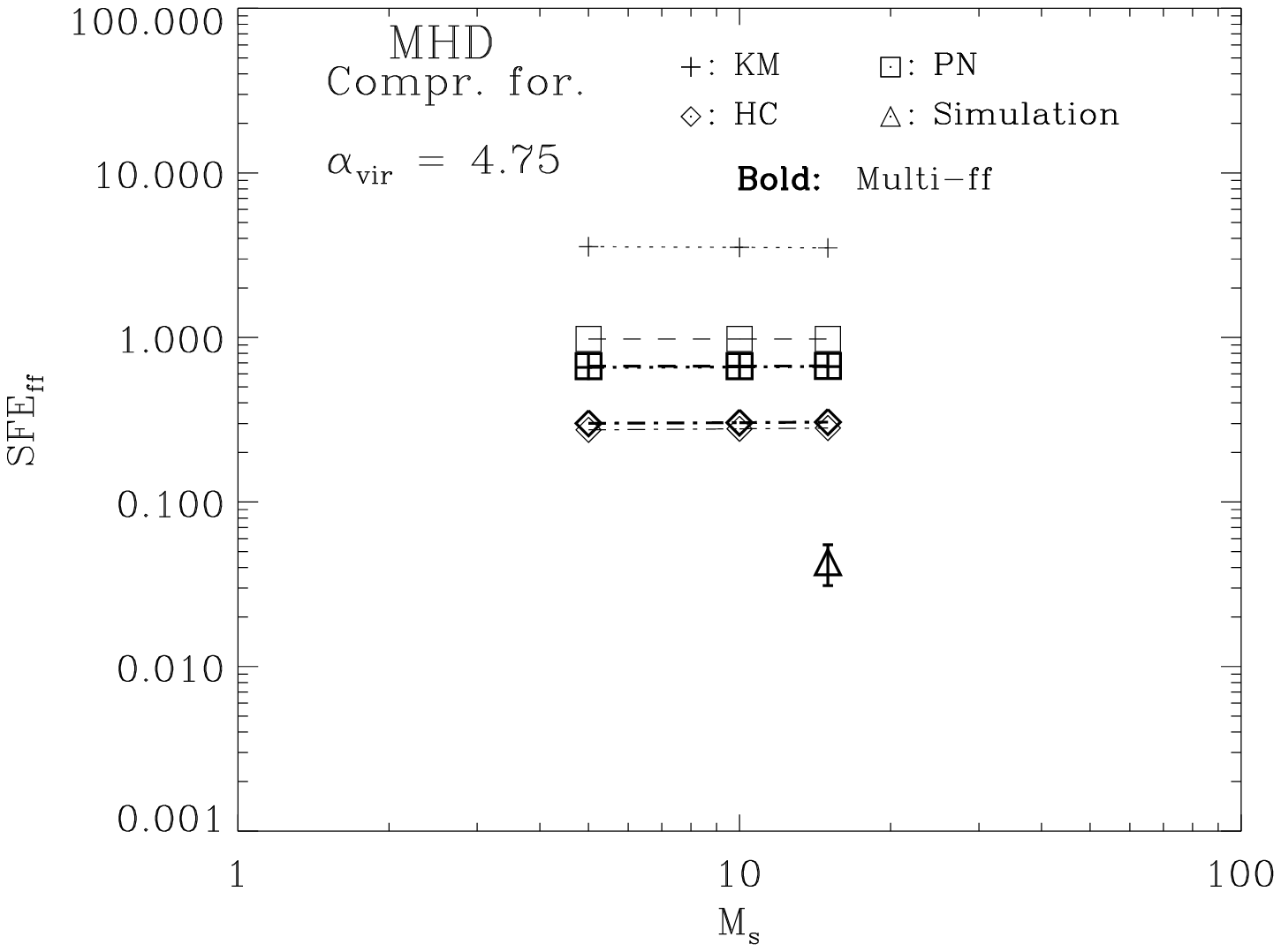}
\caption{{\it Left-hand panel:}
$\sfeff$ for the various models assuming a fully compressible forcing
($b=1$) and $\alpha =0.475$, and the $\sfeff$ for run Ms15J6C-512, with
its associated 3$\sigma$ uncertainty. {\it Right-hand panel:} same as the
middle panel, but with the model predictions calculated assuming $\alpha
=4.75$.}
\label{fig:sfeff_comp}
\end{figure*}

Finally, Fig.\ \ref{fig:accr_hist_comp} shows the mass
accretion histories for the three compressible runs, and least-squares
fits to them, with their associated slopes and uncertainties. Fig.\
\ref{fig:sfeff_comp} shows the predictions of the
various SFR models assuming fully compressible driving ($b=1$) and
virial parameter $\alpha = 0.475$ (middle) or $\alpha=4.75$ (right). As
in the case for the solenoidal run Ms15J6, it is seen again that the
models overpredict the $\sfeff$ produced by our models, even when
$\alpha$ is multiplied by a factor of 10, to mimic the larger values
obtained directly by the simulations by FK12.

We conclude from this section that the nature of the driving (solenoidal
or compressible) does not introduce any significant changes to our
results. However, we cannot rule out that, at higher resolution,
cores that are simultaneously subsonic and super-Jeans may appear.

\section{Discussion}
\label{sec:discussion} 

\subsection{Comparison with previous work}
\label{sec:mag_non-mag} 

\subsubsection{Absence of simultaneously subsonic and super-Jeans
structures, and the turbulent driving}
\label{sec:disc_subson_superJ}

Our result that no simultaneously subsonic and super-Jeans regions
(either arbitrary subboxes of the simulations or dense cores) seem to
appear in the simulations is qualitatively identical to our non-magnetic
results from Paper I. In that paper, we speculated that the
presence of the magnetic field might allow for the formation of subsonic,
super-Jeans structures because of the `cushioning' effect of the
magnetic field, which might reduce the velocity difference between the
converging flows that form the clumps. However, the fact that the
absence of these structures persists suggests that, perhaps, the
cushioning simultaneously causes the clumps to attain lower peak
densities, with both effects tending to cancel each other out.


The absence of simultaneously subsonic and super-Jeans structures in our
isothermal, driven-turbulence, magnetized simulations, in spite of their
ubiquitous observed existence in actual molecular clouds, suggests that
perhaps some other feature of our simulations is not sufficiently
realistic. One possible candidate is the very nature of the velocity
field in our simulations, which consists of randomly driven supersonic
turbulence, in which the clumps are in fact the density fluctuations
produced by the supersonic compressions in this turbulent regime, and
the driving is applied at the largest scales in the numerical box. This
continuous-driving setup is intended to model the turbulence
driven into clouds and their substructure mostly by supernova explosions
in the ambient ISM. Another important candidate is the isothermal
effective equation of state, which is one of the assumptions we are
testing, but which indeed may not be sufficiently realistic, since MCs
may well contain a mixture of atomic and molecular gas phases
\citep[e.g.,][]{LG03}.

Within this context, our three simulations, with their successively
smaller physical scales, larger mean densities, and respective forcings
applied at the largest scales available within each one, are intended to
represent a hierarchy of nested turbulent density fluctuations. However,
as already pointed out in Paper I, the equivalence between each run and
a turbulent density fluctuation of the same size within a larger scale
run is not perfect. While in each simulation the driving produces a zero
net velocity divergence, Fig.\ \ref{div1} shows that this is not the
case in high-density regions within run Ms15J6 that have the same size
as run Ms5J2, as those regions tend to have, on average, a net velocity
convergence. The same situation was encountered in Paper I.  Moreover, it
has recently been argued that the motions in numerical simulations of
molecular cloud formation, even in the presence of stellar feedback, are
dominated by hierarchical gravitational contraction \citep[i.e. of
collapses within collapses;][]{Hoyle53, Field+08, VS+09}, rather than by
random, isotropic turbulence, again
suggesting that the motions in the clouds are not equivalent to the
random, supersonic turbulent we use in the present simulations, in spite
of it being a standard procedure \citep[for a recent example,
see][]{FK12}. Moreover, there have been suggestions that this kind
of simulations may produce cores that are
systematically more dynamic than observed \citep{ABI09}. In a future
study, we plan to search for such structures in numerical simulations of
hierarchical, chaotic gravitational contraction and fragmentation
\citep[e.g.,] []{VS+07, VS+11, HH08, Hennebelle+08, Banerjee+09,
Heitsch+09, Clark+12}.

\subsubsection{The star formation efficiency per free-fall time}
\label{sec:disc_SFE}

The values of the $\sfeff$ obtained in Sec.\ \ref{sec:SFE}, including
the absence of collapse in run Ms5J2, can be compared with the
non-magnetic results from Paper I. The values of $\sfeff$ we obtained, of
0.002 for run Ms10J4 and of 0.034 for Ms15J6, are significantly smaller
than those obtained for the corresponding runs in Paper I for the
non-magnetic case, which were $\sim 0.11$ and $\sim 0.12$, respectively.
The $\sfeff$ for the present simulations, as well as for the
non-magnetic runs from Paper I, are plotted as a function of $\Ms$ in
both panels of Fig.\ \ref{fig:sfeff}.

The strong reduction in the $\sfeff$ observed in the magnetic runs
indicates that the presence of a moderately supercritical magnetic field
is able to strongly reduce the $\sfeff$ when the magneto-thermal Jeans
parameter (cf.\ Sec.\ \ref{sec:run_Ms5J2}) is sufficiently small, in
agreement with previous results \citep[e.g.,][] {Passot+95, Ostriker+99,
VS+05, NL05, PB08, FK12, FK13}. The
most dramatic difference occurs in the case of run Ms5J2, which in the
magnetic case did not produce any collapse, at least for over four
free-fall times (cf.\ Fig.\ \ref{rhomax}). Instead, the corresponding
non-magnetic run from Paper I had in fact the largest $\sfeff$ of all
runs in that paper. This implies that the {\it combined} thermal and
magnetic support must be considered in order to determine the stability
of a region since, as mentioned in Sec.\ \ref{sec:run_Ms5J2}, run M5J2
is diagnosed to be unstable by either the Jeans or the mass-to-flux
criteria taken separately.

The $\sfeff$ of our simulations can also be compared to the predictions
from the theories by KM05, PN11, and HC11. A useful summary of the
predictions from these theories, as well as simple extensions to the
magnetic case, has been recently given by
FK12. Specifically, the expressions for $\sfeff$ as a function of the
virial parameter $\alpha$, the rms Mach number, $\Ms$, the {\it plasma
beta}, $\beta$, and the {\it forcing parameter}, $b$ (which
parametrizes the relative strength of compressible and solenoidal
forcing) according to the three theories and to the multi-free-fall
variant of each one, are given in Table 1 and equations 4, 31, and
39 of FK12. The assumed values of various parameters are described in
sec.\ 2.5 of that paper, and we use the best-fitting values given in their
table 3.

Some other parameters are characteristic of our simulations.
Specifically, we use the nominal values of $\Ms$ and $\beta$ for each
run as indicated in our Table \ref{tab:run_parameters}, while, as
explained in Sec.\ \ref{sec:num_sim}, all of our runs have a nominal
value of $\alpha \approx 0.475$. Finally, since our main simulations use
purely solenoidal forcing, we take $b = 1/3$ \citep{Federrath+08}.
It is important to note that FK12 warn that their proposed extension of
the six theories to the magnetic case is applicable only for
super-Alv\'enic flows, with $\Ma \ga 2$. Since our simulations all have
$\Ma \approx 1$, they are slightly outside of the applicability range,
and thus, moderate deviations are to be expected.

The predictions from all three theories, in both their original and
`multi-free-fall' modes (a total of six theoretical models), according
to the expressions provided by FK12, are also plotted in Fig.\
\ref{fig:sfeff}, together with the values of the $\sfeff$ derived from
the simulations (cf.\ Sec.\ \ref{sec:SFE}). The {\it left panel-hand} shows
the results for the six theories in the non-magnetic (or `hydro')
case ($\beta \rightarrow \infty$) and for the simulations from Paper I,
while the {\it right-hand panel} shows the magnetic (or MHD) case, for the
values of $\beta$ corresponding to our runs (Table
\ref{tab:run_parameters}). It is seen that, in general, both the
magnitude of the $\sfeff$ and its trend with $\Ms$ is missed by the
theoretical predictions. In the non-magnetic case, only the HC11 theory,
in both its original and multi-free-fall forms, matches the magnitude of the
simulation $\sfeff$ for $\Ms =5$, but not for $\Ms=10$ and 15, since it
predicts an increasing trend with $\Ms$, while the simulations exhibit a
globally decreasing trend. This trend is only predicted by the original
KM05 theory which, however, is off in magnitude by nearly an order of
magnitude. All other theories, including the multi-free-fall KM05 one,
exhibit increasing trends with $\Ms$.

In the magnetic case, we see that none of the theories, as modified by
FK12 to include magnetic pressure, match the results from the
simulations, neither in absolute magnitude of the $\sfeff$ nor in the
trend with $\Ms$. In particular, while our simulations show a strong
trend towards collapse suppression at lower $\Ms$ with fixed $\alpha$,
the theoretical models all tend to remain at roughly constant
$\sfeff$. This suggests that the procedure used by FK12 to include the
magnetic field, consisting in simply substituting the sound speed by the
sum in quadrature of the sound and the Alfv\'en speeds to determine the
width of the density PDF and the effective Jeans length, does not
capture the effect of the decrease of $\Jeff$ at low $\Ms$ exhibited by
our simulations.

This conclusion, however, cannot be considered as definitive since, as
mentioned above, the Alv\'enic Mach number of our simulations, $\Ma
\approx 1$, is slightly outside the range of applicability
of the magnetically extended theories, as stated by FK12 ($\Ma \ga
2$). However, it appears unlikely that a difference by a mere factor of
2 in the Alfv\'enic Mach number will change the behaviour in as drastic a
manner as to go from complete collapse suppression to an independence of
$\sfeff$ from $\Ms$, as shown in the {\it right-hand panel} of Fig.\
\ref{fig:sfeff}. Moreover, note that the values we have chosen
for our parameters attempt to mimic the values observed in real regions
of the size scales represented by the simulations. If anything, it can
be argued that the magnetic field strength is somewhat excessive in the
case of the larger scale simulations (Ms10J4 and Ms15J6). However, the
chosen magnetic field strength appears to be typical for a region of
mean density $n_0 \sim 2000 \pcc$ \citep[see, e.g.] [] {Crutcher+10}, as
is the case of run Ms5J2. Therefore, run Ms5J2 may be representative of
real regions with the same physical conditions, even if it does not fall
on the range of the applicability of the magnetically extended theories.
In any case, further testing appears necessary in order to more
precisely attest the accuracy of the theoretical models and their range
of applicability.


\subsection{Implications} \label{sec:implications}

Our results have important implications for our understanding of the
role of turbulence in the support of, and regulation of SF
in, molecular clouds. Our simulations have been set up to represent
clouds obeying Larson's (1981) linewidth--size and density--size scaling
relations, so that all three of them have the same value of the virial
parameter $\alpha$, and with the kinetic energy corresponding mostly to
turbulent motions that counteract the cloud's self gravity. For such a
sequence of clouds, \citet{KT07} have argued that, for a variety of
molecular objects, the $\sfeff$ is approximately constant, at a value
$\sfeff \sim$ 2\%, independently of density, and thus, for clouds
obeying both of equations (\ref{eq:dv-r}) and (\ref{eq:n-r}),
independently of Mach number as well. Note, however, that this conclusion by
\citet{KT07} is inconsistent with equation 30 of KM05, which is a fit to
the numerical results from their theory and predicts, at constant
$\alpha$, a scaling $\sfeff \propto \Ms^{-0.32}$, as also pointed out in
section 3.2 of \citet{Elm07}.

In Paper I, we showed that our non-magnetic simulations were marginally
consistent, within their uncertainties, with the $\sfeff \sim
\Ms^{-0.32}$ dependence given by KM05. However, our magnetic simulations
from this paper suggest that this dependence is strongly modified
upon the introduction of a constant magnetic field strength, and in a
way that does not agree with KM12's extension of the KM05 theory to the
magnetic case, nor with the other two theories, in neither of their
variants. This disagreement may be attributed to the fact that the
Alfv\'enic Mach number of our simulations ($\approx 1$ in all three
runs) does not fall in the range where the extension to the magnetic
case proposed by FK12 applies. On the other hand, since the required
increment in our values of $\Ma$ to fall in the applicability range is
of only a factor of a few, it does not appear likely that the drastic
observed discrepancy can be attributed to this.


\section{Summary and conclusions} \label{sec:conclusions}

In this paper, we have presented numerical simulations of randomly
driven, supersonic, magnetized, and isothermal turbulent flows, commonly
believed to represent the flow within molecular clouds. Our simulations
are the magnetized counterparts of the simulations presented in Paper I,
and have allowed us, among other things, to determine whether, and to
what extent, the conclusions reached in that paper extend to the
magnetized case. Our main conclusions are as follows.

\begin{itemize}

\item We have used numerical simulations of continuously driven,
isothermal turbulence as a sort of `reductio ad absurdum' test of
these hypotheses, showing that they lead to results that are
inconsistent with the hypotheses. In particular, using random turbulent
driving in a box causes the clumps to have a non-random flow, but rather
having a net convergent component, so that the clumps cannot be modelled
by a simple rescaled box with random driving.

\item As in Paper I, we do not find any simultaneously subsonic and
super-Jeans structures (neither regular subboxes of the numerical box
nor dense clumps) in our simulations. In Paper I, we argued that perhaps
our failure there to find such structures was due to the neglect of the
magnetic field there, but the fact that we do not find them even in the
magnetized case strongly suggests that they form only very rarely, if at
all, in the kind of flows that we have simulated here, i.e.,
continuously and randomly driven, strongly supersonic, isothermal
turbulent flow. On the other hand, since such structures are routinely
observed in real molecular clouds \citep[see, e.g.][] {Goodman+98,
Caselli+02, ABI09}, our result suggests that this type of flow may not
be representative of the flow within molecular clouds. A viable
alternative is the type of hierarchical, chaotic gravitational
fragmentation, consisting of collapses within collapses, and
seeded by turbulence, that has been discussed in other studies
\citep[e.g.][] {CB05, VS+09, Heitsch+09, BP+11a}, to which we plan to
apply the same tests in a future study. Another viable alternative
is that, rather than being isothermal, the flow in MCs may be thermally
bistable, containing a mixture of atomic and molecular gas that may aid
in the formation of these structures.

\item Also as in Paper I, we have found that the turbulent density
enhancements (`clumps') tend to have a net negative velocity
divergence, although the typical magnitude of the convergence at a given
overdensity is decreased with respect to the non-magnetic case.
Nevertheless, the fact that this result persists indicates that a
turbulent box with totally random turbulent velocity field (thus with a
zero mean divergence) is not an exact match for the type of flow that
develops inside the clumps, which contain a non-zero net velocity
convergence. 

A crucial implication of the presence of a net convergent component
in the non-thermal motion within the clumps is that the energy contained
in these motions is not available for support against gravity, a fact
which needs to be accounted for in theories relying on this support.

\item Contrary to its non-magnetic counterpart in Paper I, which had the
highest $\sfeff$, run Ms5J2 did
not produce any collapsing objects over more than four global
free-fall times. We attributed this result to the fact that this run had
the lowest value of the magneto-thermal Jeans parameter, $\Jeff$, among
our simulations. Instead, run Ms15J6, which only had a value $\Jeff\sim
12\%$ larger than run Ms5J2, did not exhibit any delay in its
development of collapsing regions compared to the non-magnetic
case. Therefore, it appears that the transition from total to
non-existent inhibition of the collapse is a very sharp function of this
parameter. Of course, we cannot rule out the possibility that
collapse will occur in run Ms5J2 after a sufficiently long time, but in
any case, it can be concluded that the inhibition of gravitational
collapse by the combined effect of thermal pressure and the magnetic
field for this run is very strong.

\item We compared the dependence of the $\sfeff$ in our simulations (measured
as the slope of the collapsed mass versus time in units of the free-fall
time) on the turbulent Mach number $\Ms$ against the predictions of the
theories by KM05, PN11, and HC11, in both their original form, and as
modified by FK12 to include the magnetic pressure (thus forming a set of six
theoretical models in total). We compared our results against the analytic
expressions provided by FK12 for each of the six models, finding that in
general they fail to reproduce both the absolute magnitudes of the
$\sfeff$ we obtain, as well as the trend with $\Ms$, in both the
magnetic and non-magnetic cases. In particular, the suppression of
collapse in run Ms5J2 is missed by all magnetic models. 

\item The failure of the theories in the magnetic case may be explained
because the Alv\'enic Mach number of our simulations, $\Ma \approx 1$,
is too small by at least a factor of 2 to fall in the range where FK12
suggest their implementation of the effects of the magnetic field into
the theories should apply. However, since the discrepancy between the
predictions of the theories and the results of our simulations are large
even at the qualitative level (complete suppression of collapse in the
smallest, densest simulation), it is possible that the failure is an
indication of a deeper problem with the theories.

\item We suggest instead that the observed discrepancies in the
magnitude of $\sfeff$ and its dependence on $\Ms$ originate from the
fact that the assumptions of the theories are not verified in the flow
realized in the simulations. First, the velocity field in subregions of
the simulation is not a scaled-down version of the flow implemented in
the simulations as a whole: while the latter is a fully random turbulent
flow with zero net convergence, in subregions of the numerical boxes,
the flow naturally develops, on average, a globally converging
topology. This contradicts the assumption in the HC11 theory that, at
each scale, turbulence provides additional support against collapse,
since the converging component of the flow we have observed implies that
at least a fraction of the kinetic energy {\it collaborates} with the
collapse, rather than impeding it. Secondly, the KM05 and PN11 theories
assume that the collapsing objects are clumps that are simultaneously
subsonic and super-Jeans, while our simulations suggest that this is not
the dominant collapse mechanism.

\end{itemize}

In conclusion, our results seem to cast doubt on the premises of
the current SF theories and on simulations of randomly
driven turbulence as accurate representations of the physical conditions
in molecular clouds and their clumps. The fact that the motions in the
clumps have a net convergent nature implies that, far from providing
support against gravity, they will collaborate with it. Of course, as
noted in Paper I, our analysis cannot discriminate between the
convergent motions being produced by turbulence or gravity, and in
reality it is likely that both agents contribute. In the near future, we
plan to apply similar tests to a different kind of flow, suggested by
numerical simulations of the formation and evolution of entire giant
molecular clouds, in which the regime that develops seems to be
dominated by gravity at all scales, and evolves through hierarchical,
chaotic gravitational fragmentation.

\section*{Acknowledgements}

We are glad to acknowledge fruitful discussions with Christoph
Federrath. AG-S\ was supported in part by CONACyT grant 102488 to EV-S. RFG\ acknowledges support from grants PAPPIT IN117708 and
IN100511-2. The numerical simulations were performed on the cluster
KanBalam of DGTIC at UNAM, and the high-performance computing cluster,
POLARIS, at the Korea Astronomy and Space Science Institute.


\appendix

\section{On the calculation of the virial parameter $\alpha$}
\label{app:alpha_param}

Instead of the estimator given by equation (\ref{eq:alpha_M/J}), which is
based on global flow parameters, FK12 have advocated using an estimator
computed from the full computational domain, given by
\beq
\alpha = \frac{\displaystyle\sum_i M_i v_i^2}{\left|\displaystyle\sum_i M_i
\phi_i\right|}, 
\label{eq:alpha_FK12}
\eeq
where $M_i$ is the mass of the $i$th cell, $v_i$ is the magnitude of
the flow velocity in that cell, and $\phi_i$ is the gravitational
potential there. FK12 argue that, because the flow develops highly
inhomogeneous, irregularly shaped, and fractal-like density structures,
this estimator better represents the potential generated by the actual
density distribution.

However, some additional considerations are important as well. First is
the issue that in a periodic box, the Poisson equation is actually
computed as \citep[see, e.g.,][]{Weinberg72, Peebles80, AL88}
\beq
\nabla^2 \phi = 4 \pi G (\rho-\rho_0),
\label{eq:poisson}
\eeq
in order to avoid the well-known Jeans' swindle; that is, the fact that
the underlying equilibrium state in the Jeans gravitational instability
analysis is only truly self-consistent when the mean density is zero.
Equation (\ref{eq:poisson}) means that, in the simulations, the
gravitational potential arises from the distribution of density {\it
fluctuations}, rather than from the full density distribution.

In practice, this means that, in the simulation, underdense regions are
characterized by positive values of the gravitational potential. In
turn, this will cause the total sum of the gravitational term to be
decreased, in fact explaining why FK12 obtained values of $\alpha$ up to
an order of magnitude larger than that obtained with the global
parameters. This is correct with respect to the simulations, although it
reminds us that the large-scale gravitational potential in the
simulations is somewhat unrealistic, especially if the simulation
intends to represent an entire cloud. Its accuracy increases if the
modelled region is intended to represent a small fraction of a much larger
volume at the same mean density, as is the case, for example, of
simulations of cloud {\it formation} within a much larger volume
\citep[e.g.,] [] {VS+07}.

Secondly, the larger values of $\alpha$ obtained with the estimator
(\ref{eq:alpha_FK12}) seem to accomplish the opposite result of FK12's
motivation to use it in the first place: the presence of large local
density enhancements should result in a more tightly bound medium, which
would in turn result in {\it smaller} values of $\alpha$,
contrary to the larger values obtained from the estimator. 

Moreover, in order to fully capture the binding of the local, dense
structures, the local $\alpha$ parameter should be computed by taking
the velocities referred to the local centres of mass of the clumps. In
other words, if the intention of an $\alpha$ estimator is to reasonably
represent local excursions to low values of $\alpha$, it must take into
account not just the local value of the gravitational energy (the
denominator), but also the local value of the kinetic energy (the
numerator). But this must be done by removing the bulk velocity of a
local density enhancement, a consideration that is not included in the
estimator (\ref{eq:alpha_FK12}), and which in practice is very difficult
to accomplish, because the average bulk velocity to subtract depends on
the size scale of the clump.

The fact that FK12 tended to find {\it larger}, rather than smaller
values of the $\alpha$ parameter when computing it directly from the
simulation means that it was dominated by the effect of the modified
Poisson equation, rather than by the presence of strong local density
enhancements. Nevertheless, this is 
indeed more representative of the actual gravitational potential in the
simulations, due to the modified form of the Poisson equation. 

As a compromise, in the body of the paper, we compute the $\alpha$
parameter from the global parameters, but also show the predictions from
the models when $\alpha$ is multiplied by a factor of 10, which is the
maximum typical enhancement found by FK12, in order to obtain an
estimate of the modification that can be expected to occur in the model
predictions due to the computation of $\alpha$ directly from the
simulations. In general, we find that the modifications are very small.




\begin{thebibliography}{99}


\bibitem[Andr{\'e} et al.(2009)]{ABI09} Andr{\'e}, P., Basu, 
S., \& Inutsuka, S.\ 2009, Structure Formation in Astrophysics. Cambridge Univ. Press, Cambridge, p. 254 


\bibitem[Alecian \& L\'eorat(1988)]{AL88} Alecian, G., \& L\'eorat, J.\
1988, \aap, 196, 1 

\bibitem[Ballesteros-Paredes et al.(2003)]{BKV03} 
Ballesteros-Paredes, J., Klessen, R.~S., 
\& V{\'a}zquez-Semadeni, E.\ 2003, \apj, 592, 188

\bibitem[Ballesteros-Paredes et al.(2011a)]{BP+11a} 
Ballesteros-Paredes, J., Hartmann, L.~W., V{\'a}zquez-Semadeni, E., 
Heitsch, F., \& Zamora-Avil{\'e}s, M.~A.\ 2011, \mnras, 411, 65 

\bibitem[Ballesteros-Paredes et al.(2011b)]{BP+11b} 
Ballesteros-Paredes, J., V{\'a}zquez-Semadeni, E., Gazol, A., et al.\ 2011, 
\mnras, 416, 1436  


\bibitem[Banerjee et al.(2009)]{Banerjee+09} Banerjee, R., 
V{\'a}zquez-Semadeni, E., Hennebelle, P., 
\& Klessen, R.~S.\ 2009, \mnras, 398, 1082 

\bibitem[Bertoldi \& McKee(1992)]{BM92} Bertoldi, F., \& McKee, C.~F.\
1992, \apj, 395, 140 

\bibitem[Blitz(1993)]{Blitz93} Blitz, L.\ 1993, Protostars and 
Planets III, 125 

\bibitem[Bonazzola et al.(1987)]{Bonazzola+87} Bonazzola, S., Heyvaerts,
J., Falgarone, E., Perault, M., \& Puget, J.~L.\ 1987, \aap, 172, 293 

\bibitem[Brunt(2003)]{Brunt03} Brunt, C.~M.\ 2003, \apj, 584, 293 

\bibitem[Brunt et al.(2009)]{Brunt+09} Brunt, C.~M., Heyer, M.~H., \&
Mac Low, M.-M.\ 2009, \aap, 504, 883 

\bibitem[Caselli \& Myers(1995)]{CM95} Caselli, P., \&
  Myers, P.~C.\ 1995, \apj, 446, 665

\bibitem[Caselli et al.(2002)]{Caselli+02} Caselli, P., Walmsley, 
C.~M., Zucconi, A., Tafalla M., Dore L., Myers P. C., 2002, \apj, 565, 331 

\bibitem[Chandrasekhar(1951)]{Chandra51} Chandrasekhar, S.\ 1951, 
Proc. R. Soc. A, 210, 26 

\bibitem[Clark \& Bonnell(2005)]{CB05} Clark, P.~C., \& Bonnell, I.~A.\
2005, \mnras, 361, 2 

\bibitem[Clark et al.(2012)]{Clark+12} Clark, P.~C., Glover, 
S.~C.~O., Klessen, R.~S., \& Bonnell, I.~A.\ 2012, \mnras, 424, 2599 

\bibitem[Crutcher et al.(2010)]{Crutcher+10} Crutcher, R.~M., 
Wandelt, B., Heiles, C., Falgarone, E., 
\& Troland, T.~H.\ 2010, \apj, 725, 466 

\bibitem[Elmegreen(2007)]{Elm07} Elmegreen, B.~G.\ 2007, 
\apj, 668, 1064 

\bibitem[Elmegreen \& Scalo(2004)]{ES04} Elmegreen, B.~G., \& Scalo, J.\
2004, \araa, 42, 211 

\bibitem[Federrath \& Klessen(2012)]{FK12} Federrath, C., \& Klessen,
R.~S.\ 2012, \apj, 761, 156 (FK12)  

\bibitem[Federrath \& Klessen(2013)]{FK13} Federrath, C., \& Klessen,
R.~S.\ 2013, \apj, 763, 51 

\bibitem[Federrath et al.(2008)]{Federrath+08} Federrath, C., 
Klessen, R.~S., \& Schmidt, W.\ 2008, \apj, 688, L79 

\bibitem[Federrath et al.(2010)]{Federrath+10} Federrath, C.,
Roman-Duval, J., Klessen, R.~S., Schmidt, W., \& Mac Low, M.-M.\ 2010,
\aap, 512, A81 

\bibitem[Field et al.(2008)]{Field+08} Field, G.~B., Blackman, 
E.~G., \& Keto, E.~R.\ 2008, \mnras, 385, 181

\bibitem[Gibson et al.(2009)]{Gibson+09} Gibson, D., Plume, R.,
  Bergin, E., Ragan, S., \& Evans, N.\ 2009, \apj, 705, 123

\bibitem[Goldsmith \& Arquilla(1985)]{GA85} Goldsmith, P.~F., \&
Arquilla, R.\ 1985, Protostars and Planets II, Univ. Arizona Press, Tucson, AZ, p. 37 

\bibitem[G{\'o}mez et al.(2007)]{Gomez+07} G{\'o}mez, G.~C., 
V{\'a}zquez-Semadeni, E., Shadmehri, M., 
\& Ballesteros-Paredes, J.\ 2007, \apj, 669, 1042 

\bibitem[Goodman et al.(1993)]{Goodman+93} Goodman, A.~A., Benson, 
P.~J., Fuller, G.~A., \& Myers, P.~C.\ 1993, \apj, 406, 528 

\bibitem[Goodman et al.(1998)]{Goodman+98} Goodman, A.~A., 
Barranco, J.~A., Wilner, D.~J., \& Heyer, M.~H.\ 1998, \apj, 504, 223 

\bibitem[Heitsch \& Hartmann(2008)]{HH08} Heitsch, F., \& Hartmann, L.\
2008, \apj, 689, 290 

\bibitem[Heitsch et al.(2009)]{Heitsch+09} Heitsch, F., 
Ballesteros-Paredes, J., \& Hartmann, L.\ 2009, \apj, 704, 1735   

\bibitem[Hennebelle \& Chabrier(2008)]{HC08} Hennebelle, P., \&
Chabrier, G.\ 2008, \apj, 684, 395 

\bibitem[Hennebelle \& Chabrier(2011)]{HC11} Hennebelle, P., \&
Chabrier, G.\ 2011, \apj, 743, L29 (HC11)

\bibitem[Hennebelle et al.(2008)]{Hennebelle+08} Hennebelle, P.,
Banerjee, R., V{\'a}zquez-Semadeni, E., Klessen, R.~S., \& Audit, E.\
2008, \aap, 486, L43


\bibitem[Heyer et al.(2009)]{Heyer+09} Heyer, M., Krawczyk, C.,
  Duval, J., \& Jackson, J.~M.\ 2009, \apj, 699, 1092


\bibitem[Hopkins(2012a)]{Hopkins12a} Hopkins, P.~F.\ 2012, \mnras, 
423, 2016 

\bibitem[Hopkins(2012b)]{Hopkins12b} Hopkins, P.~F.\ 2012, \mnras, 
423, 2037 

\bibitem[Hopkins(2013)]{Hopkins13} Hopkins, P.~F.\ 2013, \mnras, 
430, 1653 

\bibitem[Hoyle(1953)]{Hoyle53} Hoyle, F.\ 1953, \apj, 118, 513 

\bibitem[Imara \& Blitz(2011)]{IB11} Imara, N., \& Blitz, L.\ 2011,
\apj, 732, 78 

\bibitem[Imara et al.(2011)]{IBB11} Imara, N., Bigiel, F., 
\& Blitz, L.\ 2011, \apj, 732, 79 

\bibitem[Kegel(1989)]{Kegel89} Kegel, W.~H.\ 1989, \aap, 225, 517 

\bibitem[Kim et al.(1999)]{Kim+99} Kim, J.,
Ryu, D., Jones, T. W., \& Hong, S. S. 1999, ApJ, 514, 506 

\bibitem[Krumholz \& McKee(2005)]{KM05} Krumholz, M.~R., \& McKee,
C.~F.\ 2005, \apj, 630, 250 (KM05) 

\bibitem[Krumholz \& Tan(2007)]{KT07} Krumholz, M.~R., \& Tan, J.~C.\
2007, \apj, 654, 304 

\bibitem[Larson(1981)]{Larson81} Larson, R. B. 1981, MNRAS, 194, 809


\bibitem[Li \& Goldsmith(2003)]{LG03} Li, D., \& Goldsmith, P.~F.\ 2003,
\apj, 585, 823 

\bibitem[Li et al.(2012)]{Li+12} Li, J., Wang, J., Gu, Q., 
Zhang, Z.-y., \& Zheng, X.\ 2012, \apj, 745, 47 

\bibitem[McKee \& Ostriker(2007)]{MO07} McKee, C.~F., \& Ostriker,
E.~C.\ 2007, \araa, 45, 565

\bibitem[Mac Low \& Klessen(2004)]{MK04} Mac Low, M.-M., \& Klessen,
R.~S.\ 2004, Rev. Mod. Phys., 76, 125  

\bibitem[Molina et al.(2012)]{Molina+12} Molina, F.~Z., Glover, 
S.~C.~O., Federrath, C., \& Klessen, R.~S.\ 2012, \mnras, 423, 2680 


\bibitem[Myers \& Goodman(1988)]{MG88} Myers, P.~C., \& Goodman, A.~A.\
1988, \apj, 326, L27 

\bibitem[Nakamura \& Li(2005)]{NL05} Nakamura, F., \& Li, Z.-Y.\ 2005,
\apj, 631, 411 

\bibitem[Nakano \& Nakamura(1978)]{NN78} Nakano, T., \& Nakamura, T.\
1978, \pasj, 30, 671 

\bibitem[Ostriker et al.(1999)]{Ostriker+99} Ostriker, E.~C., 
Gammie, C.~F., \& Stone, J.~M.\ 1999, \apj, 513, 259 

\bibitem[Padoan(1995)]{Padoan95} Padoan, P.\ 1995, \mnras, 277, 
377 


\bibitem[Padoan \& Nordlund(2011)]{PN11} Padoan, P., \& Nordlund,
{\AA}.\ 2011, \apj, 730, 40 (PN11)

\bibitem[Padoan et al.(1997)]{Padoan+97} Padoan, P., Nordlund, 
A., \& Jones, B.~J.~T.\ 1997, \mnras, 288, 145 

\bibitem[Passot \& V{\'a}zquez-Semadeni(1998)]{PV98} Passot, T., \&
V{\'a}zquez-Semadeni, E.\ 1998, \pre, 58, 4501 

\bibitem[Passot et al.(1995)]{Passot+95} Passot, T., 
Vazquez-Semadeni, E., \& Pouquet, A.\ 1995, \apj, 455, 536 

\bibitem[Peebles(1980)]{Peebles80} Peebles, P.~J.~E.\ 1980, 
Research supported by the National Science Foundation. 
Princeton University Press,~Princeton, NJ,p.~435  

\bibitem[Phillips(1999)]{Phillips99} Phillips, J.~P.\ 1999, \aaps, 134, 241 

\bibitem[Plume et al.(1997)]{Plume+97} Plume, R., Jaffe, D.~T.,
  Evans, N.~J., II, Martin-Pintado, J., \& Gomez-Gonzalez, J.\ 1997,
  \apj, 476, 730

\bibitem[Price \& Bate(2008)]{PB08} Price, D.~J., \& Bate, M.~R.\ 2008,
\mnras, 385, 1820 

\bibitem[Robertson \& Goldreich(2012)]{RG12} Robertson, B., \&
Goldreich, P.\ 2012, \apj, 750, L31  

\bibitem[Rosolowsky(2007)]{Rosol07} Rosolowsky, E.\ 2007, \apj, 
654, 240 


\bibitem[Scalo(1990)]{Scalo90} Scalo, J.\ 1990, in Capuzzo-Dolcetta R., Chiosi C., di Fazio A., eds, Astrophysics and Space Science Library, Vol. 162, Physical Processes in
  Fragmentation and Star Formation. Springer-Verlag, Berlin, p. 151

\bibitem[Shirley et al.(2003)]{Shirley+03} Shirley, Y.~L., Evans,
  N.~J., II, Young, K.~E., Knez, C., \& Jaffe, D.~T.\ 2003, \apjs,
  149, 375

\bibitem[Shu(1992)]{Shu92} Shu, F.~H.\ 1992, Physics of 
Astrophysics, Vol.~II. University Science 
Books, Mill Valley, CA

\bibitem[Truelove et al.(1997)]{Truelove+97} Truelove, J.~K., 
Klein, R.~I., McKee, C.~F., Holliman J. H., II, Howell., Greenough J. A., 1997, \apj, 489, L179 

\bibitem[V\'azquez-Semadeni(1994)]{VS94} V\'azquez-Semadeni, E.\ 
1994, \apj, 423, 681 

\bibitem[V\'azquez-Semadeni \& Gazol(1995)]{VG95} V\'azquez-Semadeni, E., \&
Gazol, A.\ 1995, \aap, 303, 204  

\bibitem[V\'azquez-Semadeni et al.(1996)]{VS+96} 
V\'azquez-Semadeni, E., Passot, T., \& Pouquet, A.\ 1996, \apj, 473, 881 

\bibitem[V\'azquez-Semadeni et al.(1998)]{VS+98} 
V\'azquez-Semadeni, E., Cant\'o, J., \& Lizano, S.\ 1998, \apj, 492, 596 

\bibitem[V\'azquez-Semadeni et al.(2000)]{VS+00} 
V\'azquez-Semadeni, E., Ostriker, E.~C., Passot, T., Gammie, C.~F., 
\& Stone, J.~M.\ 2000, Protostars and Planets IV. Univ. Arizona Press, Tucson, AZ, p. 3

\bibitem[V{\'a}zquez-Semadeni et al.(2003)]{VBK03} 
V{\'a}zquez-Semadeni, E., Ballesteros-Paredes, J., 
\& Klessen, R.~S.\ 2003, \apj, 585, L131 

\bibitem[V{\'a}zquez-Semadeni et al.(2005)]{VS+05} 
V{\'a}zquez-Semadeni, E., Kim, J., Shadmehri, M., 
\& Ballesteros-Paredes, J.\ 2005, \apj, 618, 344  

\bibitem[V{\'a}zquez-Semadeni et al.(2007)]{VS+07} 
V{\'a}zquez-Semadeni, E., G{\'o}mez, G.~C., Jappsen, A.~K., Ballesteros-Paredes J., Gonz{\'a}lez R. F., Klessen R. S., 2007, 
\apj, 657, 870 

\bibitem[V\'azquez-Semadeni et al.(2008)]{VS+08}
  V\'azquez-Semadeni, E., Gonz\'alez, R. F., Ballesteros-Paredes, J.,
  Gazol, A., \& Kim, J., 2008,   MNRAS, 390, 769 (Paper I)

\bibitem[V{\'a}zquez-Semadeni et al.(2009)]{VS+09} 
V{\'a}zquez-Semadeni, E., G{\'o}mez, G.~C., Jappsen, A.-K., 
Ballesteros-Paredes, J., \& Klessen, R.~S.\ 2009, \apj, 707, 1023


\bibitem[V{\'a}zquez-Semadeni et al.(2011)]{VS+11} V\'azquez-Semadeni,
E., Banerjee, R., G\'omez G. C., et al. \ 2011, MNRAS, 414, 2511 


\bibitem[Weinberg(1972)]{Weinberg72} Weinberg, S.\ 1972, 
Gravitation and Cosmology: Principles and Applications of the General 
Theory of Relativity. Wiley New York, pp. 688, (ISBN 
0-471-92567-5)  

\bibitem[Wu et al.(2010)]{Wu+10} Wu, J., Evans, N.~J., 
Shirley, Y.~L., \& Knez, C.\ 2010, \apjs, 188, 313 

\bibitem[Zamora-Avil{\'e}s et al.(2012)]{ZVC12} Zamora-Avil{\'e}s, M.,
V{\'a}zquez-Semadeni, E., \& Col{\'{\i}}n, P.\ 2012, \apj, 751, 77 

\bibitem[Zuckerman \& Evans(1974)]{ZE74} 
Zuckerman, B., \& Evans, N. J. 1974, ApJ, 192, L149

\end{thebibliography}
\end{document}